\PassOptionsToPackage{unicode}{hyperref}
\PassOptionsToPackage{hyphens}{url}
\PassOptionsToPackage{dvipsnames,svgnames,x11names}{xcolor}
\documentclass[
]{article}
\usepackage{xcolor}
\usepackage{hyperref}
\usepackage[margin=1in]{geometry}
\usepackage{amsmath,amssymb}
\usepackage{lineno}
\usepackage{iftex}
\ifPDFTeX
  \usepackage[T1]{fontenc}
  \usepackage[utf8]{inputenc}
  \usepackage{textcomp} 
\else 
  \usepackage{unicode-math} 
  \defaultfontfeatures{Scale=MatchLowercase}
  \defaultfontfeatures[\rmfamily]{Ligatures=TeX,Scale=1}
\fi
\usepackage{lmodern}
\ifPDFTeX\else
\fi
\IfFileExists{upquote.sty}{\usepackage{upquote}}{}
\IfFileExists{microtype.sty}{
  \usepackage[]{microtype}
  \UseMicrotypeSet[protrusion]{basicmath} 
}{}
\makeatletter
\@ifundefined{KOMAClassName}{
  \IfFileExists{parskip.sty}{%
    \usepackage{parskip}
  }{
    \setlength{\parindent}{0pt}
    \setlength{\parskip}{6pt plus 2pt minus 1pt}}
}{
  \KOMAoptions{parskip=half}}
\makeatother
\makeatletter
\ifx\paragraph\undefined\else
  \let\oldparagraph\paragraph
  \renewcommand{\paragraph}{
    \@ifstar
      \xxxParagraphStar
      \xxxParagraphNoStar
  }
  \newcommand{\xxxParagraphStar}[1]{\oldparagraph*{#1}\mbox{}}
  \newcommand{\xxxParagraphNoStar}[1]{\oldparagraph{#1}\mbox{}}
\fi
\ifx\subparagraph\undefined\else
  \let\oldsubparagraph\subparagraph
  \renewcommand{\subparagraph}{
    \@ifstar
      \xxxSubParagraphStar
      \xxxSubParagraphNoStar
  }
  \newcommand{\xxxSubParagraphStar}[1]{\oldsubparagraph*{#1}\mbox{}}
  \newcommand{\xxxSubParagraphNoStar}[1]{\oldsubparagraph{#1}\mbox{}}
\fi
\makeatother

\usepackage{longtable,booktabs,array}
\usepackage{calc} 
\usepackage{etoolbox}
\makeatletter
\patchcmd\longtable{\par}{\if@noskipsec\mbox{}\fi\par}{}{}
\makeatother
\IfFileExists{footnotehyper.sty}{\usepackage{footnotehyper}}{\usepackage{footnote}}
\makesavenoteenv{longtable}
\usepackage{graphicx}
\makeatletter
\newsavebox\pandoc@box
\newcommand*\pandocbounded[1]{
  \sbox\pandoc@box{#1}%
  \Gscale@div\@tempa{\textheight}{\dimexpr\ht\pandoc@box+\dp\pandoc@box\relax}%
  \Gscale@div\@tempb{\linewidth}{\wd\pandoc@box}%
  \ifdim\@tempb\p@<\@tempa\p@\let\@tempa\@tempb\fi
  \ifdim\@tempa\p@<\p@\scalebox{\@tempa}{\usebox\pandoc@box}%
  \else\usebox{\pandoc@box}%
  \fi%
}
\def\fps@figure{htbp}
\makeatother

\usepackage[]{natbib}
\usepackage{amssymb}
\usepackage{booktabs}
\usepackage{authblk}
\usepackage{rotating}
\makeatletter
\@ifpackageloaded{caption}{}{\usepackage{caption}}
\AtBeginDocument{%
\ifdefined\contentsname
  \renewcommand*\contentsname{Table of contents}
\else
  \newcommand\contentsname{Table of contents}
\fi
\ifdefined\listfigurename
  \renewcommand*\listfigurename{List of Figures}
\else
  \newcommand\listfigurename{List of Figures}
\fi
\ifdefined\listtablename
  \renewcommand*\listtablename{List of Tables}
\else
  \newcommand\listtablename{List of Tables}
\fi
\ifdefined\figurename
  \renewcommand*\figurename{Figure}
\else
  \newcommand\figurename{Figure}
\fi
\ifdefined\tablename
  \renewcommand*\tablename{Table}
\else
  \newcommand\tablename{Table}
\fi
}
\@ifpackageloaded{float}{}{\usepackage{float}}
\floatstyle{ruled}
\@ifundefined{c@chapter}{\newfloat{codelisting}{h}{lop}}{\newfloat{codelisting}{h}{lop}[chapter]}
\floatname{codelisting}{Listing}

\makeatother
\makeatletter
\@ifpackageloaded{caption}{}{\usepackage{caption}}
\@ifpackageloaded{subcaption}{}{\usepackage{subcaption}}
\makeatother
\usepackage{bookmark}
\IfFileExists{xurl.sty}{\usepackage{xurl}}{} 
\hypersetup{
  pdftitle={Regime-Aware Estimation of Large Equity Correlation Matrices: Tail-Pairwise Dependence and Dynamic Conditional Correlation Decomposition for Water and Energy Sectors},
  pdfauthor={Ensor, Katherine B.; Popoola, Quadri; Kruse, Meridith; Cooley, Danial S.},
  colorlinks=true,
  linkcolor={blue},
  filecolor={Maroon},
  citecolor={Blue},
  urlcolor={Blue},
  pdfcreator={LaTeX via pandoc}}

\title{Regime-Aware Estimation of Large Equity
Correlation Matrices: Tail-Pairwise Dependence and Dynamic Conditional Correlation Decomposition for Water
and Energy Sectors}
\author[1]{Ensor, Katherine B.*} 
\author[2]{Popoola, Quadri}
\author[2]{Kruse, Meredith}
\author[3]{Cooley, Daniel S.}
\affil[1]{*Corresponding author, ensor@rice.edu, Center for Computational Finance and Economic Systems, Department of Statistics, Rice University, Houston, TX, U.S.A.}
\affil[2]{Center for Computational Finance and Economic Systems, Department of Statistics, Rice University, Houston, TX, U.S.A.}
\affil[3]{Department of Statistics, Colorado State University, Fort Collins, CO, U.S.A.}

\date{\today}
\begin{document}

\maketitle

\section{Abstract}\label{abstract}

Equity correlation matrices are foundational to quantitative risk, yet the operational default of a single Gaussian-copula structure fit to the full sample has a structural defect: it forces asymptotic tail-independence and so cannot represent the joint moves that drive risk in stress episodes. We argue that the dependence of equity returns is not one object but three distinct under lower-tail, central, and upper-tail conditions, and that any estimator aimed at tail-sensitive applications should be regime-aware. We develop a decomposition that fits a dedicated dependence matrix to each regime, using a tail-dependence estimator grounded in multivariate regular variation in the two tails and a time-varying conditional-correlation model in the central regime, and recombines the three at the unconditional regime probabilities to form a steady-state matrix. The construction identifies tail-regime dependence that the Gaussian-copula benchmark structurally suppresses. Demonstrated on water- and energy-sector Exchange Traded Funds (ETF) constituents, the tail regimes reorganize the commodity partition produced by the central regime, identifying the pairs where the Gaussian-copula default systematically fails to identify joint extreme co-movement.

\section{Introduction}\label{introduction}

The estimation of large correlation and covariance matrices is a foundational task in financial risk management, underpinning portfolio construction, Value-at-Risk calculations, stress testing, and derivatives pricing. For practical applications a single point estimate is rarely sufficient: portfolios contain hundreds or thousands of names, dependence structure evolves with the macro-financial cycle, and the joint behavior of returns during stress episodes routinely departs from
the Gaussian ideal. These three pressures (scale, non-stationarity, and tail asymmetry) motivate a methodology that is simultaneously computationally tractable, time-varying, and explicitly regime-aware. For a broader review of the computational-finance landscape on correlation, volatility, and markets, see \citet{EnsorKoev2014}. To understand the importance of tail events and a novel Bayesian method of modeling, we recommend \citet{VankovEtAl2019}.

The substantial literature documents that dependence in equity returns is neither stationary nor symmetric across the return distribution. \citet{Engle2002} and \citet{TseTsui2002} introduced Dynamic Conditional Correlation (DCC) models that allow the correlation matrix to evolve over time while factoring the multivariate problem into tractable univariate GARCH stages. \citet{CappielloEtAl2006} extended the framework to allow downside shocks to drive larger correlation responses than upside shocks of comparable magnitude, a feature strongly supported in the equity data. The DCC family has continued to evolve along several axes since then. \citet{BauwensXu2023} augment the standard specification with realized variances and correlations built from intraday data, and \citet{ArchakovEtAl2025} develop a fully multivariate Realized GARCH model, both directions that strengthen the univariate filter on which the multivariate stage rests. A simultaneous literature has worked to scale multivariate GARCH models to high cross-sections through penalized estimation \citep{YaoZouXing2024} and explicit cluster structure in the
conditional correlation and tail-dependence matrices \citep{TongHansenArchakov2025}. In parallel, the literature on multivariate extremes, in particular the Tail Pairwise Dependence Matrix of \citet{CooleyThibaud2019}, has provided an alternative to copula-based tail models (see also \citet{yuen202070, mhatre2024transformed}). The TPDM summarizes pairwise tail co-movement through a moment-type estimator defined on data transformed to common regularly-varying margins, and unlike the Gaussian copula, it neither imposes tail symmetry nor requires a parametric family. Recent work has situated estimators of this kind within the broader theory of graphical models for extremes; \citet{EngelkeEtAl2024} provide a comprehensive review of this rapidly developing area, which provides parsimonious statistical tools for extremal dependence in high-dimensional problems.

In this paper, we combine these two threads. We partition the return history into three regimes (lower tail, time-varying central, and upper tail) defined by univariate excursions of the standardized residuals across the cross-section of assets, and we fit a dedicated dependence estimator in each. For the two tails, we estimate TPDMs after a Fréchet (inverse-\(\alpha\)) margin transformation, following the Cooley and Thibaud construction. These estimates are not time dependent, but rather identify the cross-sectional relationships spanning that remain after the univariate return series are adjusted for temporal GARCH structure.  For the central regime, we fit a symmetric DCC-GARCH \citep{Engle2002} with a multivariate Gaussian likelihood, the standard benchmark specification of the DCC family. The three regime-specific matrices are then combined into a steady-state (long-run) dependence matrix via a convex combination whose weights reflect the unconditional probability of being in each regime. This framework facilitates a more comprehensive characterization of the typical time-varying correlation structure among assets, as well as the systematic deviations of the typical structure observed during periods of market contraction and expansion.

We take as our empirical setting the constituents of two commodity-benchmark ETFs, the Invesco Water Resources ETF (CGW) and the Energy Select Sector SPDR (XLE), giving a panel that is economically interpretable and that spans two commodities, represented by underlying assets within the ETFs, with meaningfully different exposure to macro shocks.  The water-energy nexus has emerged as a coherent investment theme in its own right, reflecting that the two commodities are linked by shared physical, regulatory, and climate-policy drivers. For example, \citet{RaathEnsor2023} develop wavelet-L2E stochastic volatility models in this setting and document time-frequency volatility structure that is specific to water and energy securities. The choice is also motivated by a recent body of empirical work documenting non-trivial tail dependence and time-varying spillover among energy- and clean-energy-related ETFs \citep{JiEtAl2024, BanerjeeEtAl2024, OzkanEtAl2024, EnsorEtAl2020} and the asymmetric tail-risk co-movement between energy markets and exogenous climate-risk factors \citep{AdeabahPham2025}, which together suggest that a regime-aware dependence estimator is well-suited to this
application context. 

The remainder of the paper proceeds as follows. Section 2 presents the methodology: the univariate GARCH(1,1) filter, the middle-region partition, the symmetric DCC for the central regime, the TPDM construction for the two tails and the steady-state convex combination. We also review standard methods of estimating correlation matrices, namely through the Gaussian copula, Pearson and Spearman correlations. As well as, our basic approaches of comparisons used. Section 3 documents the data and preprocessing pipeline, which includes a Winsorization step before estimating the central regime time-varying dynamics. Section 4 presents the empirical results: univariate fits, the conditional-correlation dynamics of the DCC, the upper- and lower-tail TPDMs, the latent factor structure across all dependence estimators. Comparisons are made for the lower, middle, and upper dynamics, the steady state correlation, Spearman and Pearson correlation matrices. The Gaussian Copula correlation estimate is used to test the significance of the dependence observed in the tails from the TPDM construction. At the end of Section 4, the computational costs are discussed. Section 5 concludes the manuscript by providing a detailed discussion of the empirical and theoretical findings and by identifying the partial tail-correlation coefficient of \citet{GongEtAl2024}, together with the inferential framework of \citet{KimLee2024}, as a natural and immediate avenue for further research. This extension would generalize the marginal TPDM-based analysis to a conditional-independence framework within an extremal graphical model setting.

\section{Methodology}\label{methodology}

\begin{figure}[H]

\centering{

\includegraphics[width=.9\textwidth,height=.85\textheight]{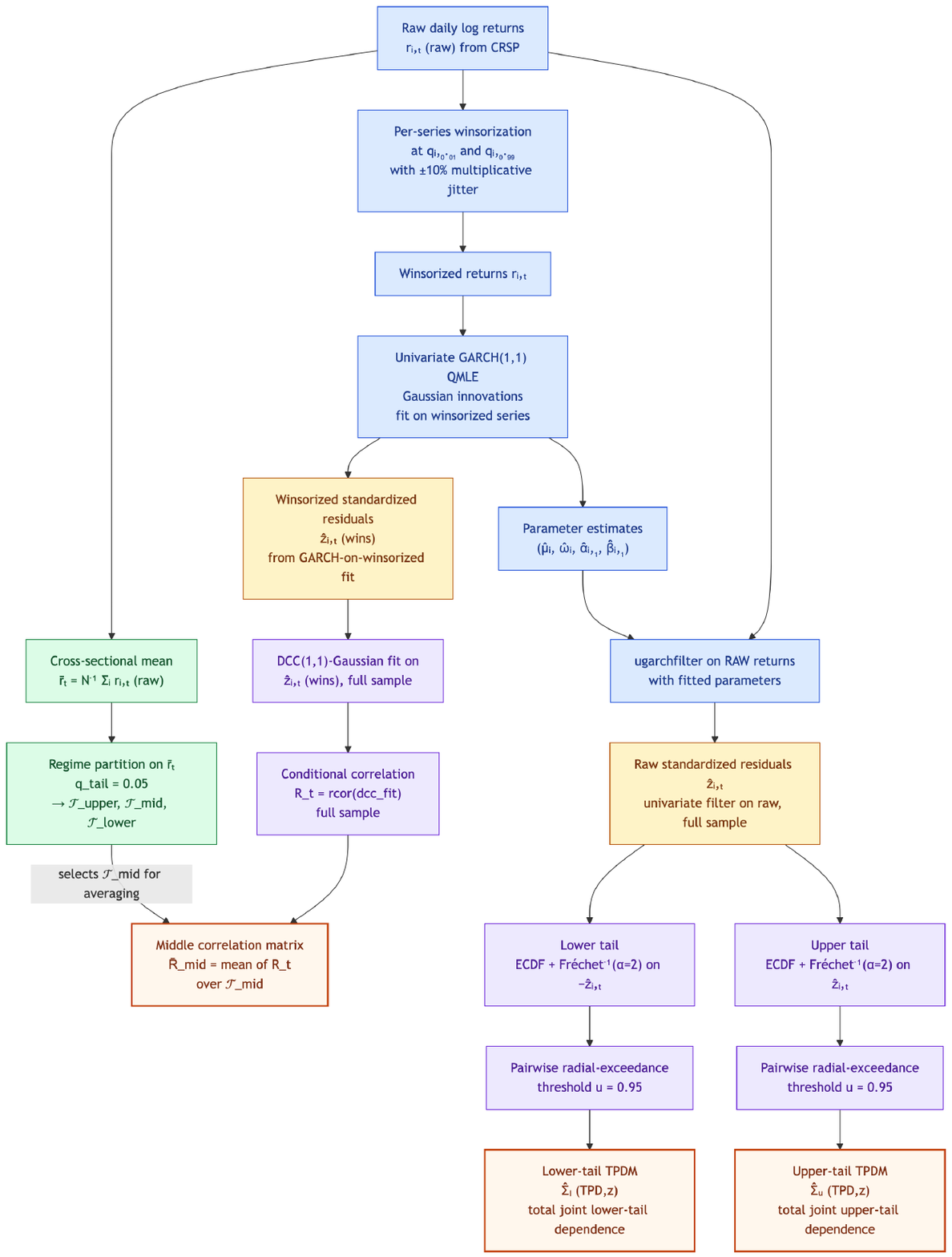}

}
\vskip24truept
\caption{\label{fig-pipeline} End-to-end estimator pipeline from raw CRSP daily log returns to compute $R_t$, $\bar R_{\text{mid}}$, \(\widehat\Sigma^{\text{TPD},z}_{l}\), \(\widehat\Sigma^{\text{TPD},z}_{u}\) (latter three are branch endpoints). Univariate filtering yields two parallel standardized-residual streams sharing one fitted parameter vector: $\widehat z^{\,\text{wins}}$ (GARCH fit on winsorized series, feeds the DCC) and $\widehat z$ (`ugarchfilter` on raw, feeds the TPDM). The DCC is fit to the full sample of the winsorized standardized residuals $\widehat z^{\,\text{wins}}$ and produces a time-varying conditional correlation $R_t$ on every date; the regime partition (green) is panel-wide (defined on the cross-sectional mean $\bar r_t$) and enters only as the selector for the date subset $\mathcal T_{\text{mid}}$ over which $R_t$ is averaged to form $\bar R_{\text{mid}}$. Downstream the steady-state dependence matrix is built from $\bar R_{\text{mid}}$ and the two tail TPDMs.}

\end{figure}

This section introduces the regime-aware estimator as a sequence of five mathematical stages: a univariate GARCH filter that generates standardized residuals; a partition of the return panel into three cross-sectional regimes; the estimation of a symmetric DCC model for the central regime; the estimation of a TPDM for each of the two tail regimes; and, finally, the construction of a single steady-state correlation matrix as a convex combination of the three regime-specific matrices. Figure~\ref{fig-pipeline} summarizes the end-to-end estimator pipeline from the raw CRSP daily log returns through the central-regime correlation matrix and the upper- and lower-tail TPDMs.

Subsequently, we describe the methodological framework employed to analyze and compare the correlation structures associated with the three regimes. To address the question of whether the tail correlations are independent we implement a simulation based test assuming Gaussian copula correlations for the null hypothesis. For general comparison of the steady-state correlation we target the standard Pearson and Spearman correlation matrices.   

\subsubsection{Returns and winsorized returns}\label{ssec:returns} Let \(r_{i,t}^{\text{raw}}\) denote the raw daily log return of asset \(i \in \{1,\dots,N\}\) at time \(t \in \{1,\dots,T\}\). As a preprocessing step we winsorize each return at the empirical \(1\%\) and \(99\%\) quantiles of its own series's return distribution, with a multiplicative \(\pm 10\%\) jitter around the boundary so the marginal distribution does not accumulate a mass spike at the quantile. Letting \(q_{i, \alpha}\) denote the empirical \(\alpha\)-quantile of asset \(i\)'s own raw return distribution, \[
r_{i,t}^{\text{wins}} \;=\; \begin{cases}
U^{+}_{i,t} \sim \mathrm{Uniform}\!\bigl(0.9\, q_{i,\,0.99},\ 1.1\, q_{i,\,0.99}\bigr), & r_{i,t}^{\text{raw}} > q_{i,\,0.99}, \\
U^{-}_{i,t} \sim \mathrm{Uniform}\!\bigl(1.1\, q_{i,\,0.01},\ 0.9\, q_{i,\,0.01}\bigr), & r_{i,t}^{\text{raw}} < q_{i,\,0.01}, \\
r_{i,t}^{\text{raw}}, & \text{otherwise.}
\end{cases}
\] 
Each series is winsorized separately because dispersion and tail heaviness vary meaningfully across the cross-section. For example, as we will see, high-volatility energy names have wider 1\% and 99\% quantile bands than low-volatility water names, so a panel-wide or even ETF-wide threshold would over-cap the calmer series and under-cap the more volatile ones. Estimating \(q_{i, 0.01}\) and \(q_{i, 0.99}\) series-by-series respects each asset's own marginal scale and replaces each extreme observation by a draw centered on the same series's tail quantile. Since \(q_{i, 0.01} < 0 < q_{i, 0.99}\) for every series in this cross-section, the sign of each capped observation is preserved automatically. In other words, lower-tail draws are negative and upper-tail draws are positive. The seed is fixed in the implementation so that the random draws are reproducible across renders. 

The purpose of winsorizing the returns is to ensure the parameter estimates in the DCC-GARCH models are not influenced by extreme returns. 

\subsection{Univariate GARCH Volatility
Filter}\label{sec-ar-garch-filter}

We model each series with a standard GARCH(1,1) specification with Gaussian innovations, 
\[ 
r_{i,t}^{\text{wins}} = \mu_i + \epsilon_{i,t}, \qquad \epsilon_{i,t} = \sigma_{i,t}\, z_{i,t}, \qquad  \sigma_{i,t}^{2} = \omega_i + \alpha_{i,1}\,\epsilon_{i,t-1}^{2} + \beta_{i,1}\,\sigma_{i,t-1}^{2}, 
\]
with \(z_{i,t} \stackrel{\text{iid}}{\sim} \mathcal{N}(0, 1)\) under the working model. The variance intercept \(\omega_i\) is estimated alongside the persistence pair \((\alpha_{i,1}, \beta_{i,1})\), and stationarity of the conditional variance is ensured by \(\alpha_{i,1} + \beta_{i,1} < 1\), which delivers the finite unconditional variance \(E[\sigma_{i,t}^{2}] = \omega_i / (1 - \alpha_{i,1} - \beta_{i,1})\). The parameter vector \((\mu_i, \omega_i, \alpha_{i,1}, \beta_{i,1})\) is estimated by quasi-MLE on the winsorized series \(\{r_{i,t}^{\text{wins}}\}_{t=1}^T\), so the parameter estimates are not driven by the tail observations that the per-series cap above has rescaled. \citet{BollerslevWooldridge1992} sandwich standard errors give heteroskedasticity-robust inference under misspecification of the Gaussian innovation distribution. 

Once the parameter vector \((\widehat\mu_i, \widehat\omega_i, \widehat\alpha_{i,1}, \widehat\beta_{i,1})\) has been estimated on the winsorized series, the GARCH(1,1) recursion is run twice to produce two parallel full-sample standardized-residual streams that share one fitted parameter vector. The first is the \emph{winsorized} residual stream, formed by the QMLE step itself on the winsorized series, 
\[ 
\widehat\sigma_{i,t}^{\,\text{wins},2} \;=\; \widehat\omega_i + \widehat\alpha_{i,1}\,(r_{i,t-1} - \widehat\mu_i)^{2} + \widehat\beta_{i,1}\,\widehat\sigma_{i,t-1}^{\,\text{wins},2}, \qquad \widehat z^{\,\text{wins}}_{i,t} \;=\; \frac{r_{i,t} - \widehat\mu_i}{\widehat\sigma^{\,\text{wins}}_{i,t}}. 
\]
The second is the \emph{raw} residual stream, obtained by applying the GARCH(1,1) recursion with the same coefficients to the original (un-winsorized) return series, 
\[
\widehat\sigma_{i,t}^{\,2} \;=\; \widehat\omega_i + \widehat\alpha_{i,1}\,(r_{i,t-1}^{\text{raw}} - \widehat\mu_i)^{2} + \widehat\beta_{i,1}\,\widehat\sigma_{i,t-1}^{\,2}, \qquad \widehat z_{i,t} \;=\; \frac{r_{i,t}^{\text{raw}} - \widehat\mu_i}{\widehat\sigma_{i,t}}. 
\] 
Both streams are defined on the full sample \(\{1,\dots,T\}\) with no missing-value gaps. They feed different downstream stages. First, the estimation of DCC parameters is run on the winsorized residual stream \(\widehat z^{\,\text{wins}}_{i,t}\), so the multivariate parameters of the central-regime are not driven by the extreme returns that the winsorization per-series has already rescaled. Second, the TPDM is applied to the raw residual stream \(\widehat z_{i,t}\), so the heavy-tail observations the radial-exceedance estimator depends on are preserved. Both residual streams are summarized graphically in Figure~\ref{fig-pipeline}.  

\subsection{Middle-Regime Partition and Dynamic Conditional
Correlation}\label{sec-middle-regime-partition}

We fit the symmetric DCC-GARCH of \citet{Engle2002} on the full-sample winsorized standardized residual stream \(\widehat z^{\,\text{wins}}_{i,t}\) produced. The conditional covariance \(H_t\) is decomposed as  
\[ 
H_t = D_t R_t D_t,\qquad R_t = \mathrm{diag}(Q_t)^{-1/2}\, Q_t\, \mathrm{diag}(Q_t)^{-1/2}, 
\]
where \(D_t\) is a diagonal matrix of conditional standard deviations from the univariate GARCH fits, and the auxiliary \(Q_t\) follows the symmetric recursion  
\[ 
Q_t = (1 - a - b)\,\bar Q + a\, z_{t-1}^{\,\text{wins}} z_{t-1}^{\text{wins}\top} + b\, Q_{t-1}, 
\]
with \(z_{t}^{\,\text{wins}}\) the vector of standardized winsorized residuals at time \(t\), \(\bar Q\) the unconditional covariance of \(z_{t}^{\,\text{wins}}\), and \((a, b)\) the DCC parameters. Setting \(a = b = 0\) recovers the constant conditional correlation model of \citet{Bollerslev1990}. The model is fit with a multivariate Gaussian likelihood at the multivariate stage. This model and fit is the standard benchmark specification of the DCC family and treats the standardized residuals as conditionally Gaussian.  

We summarize the central regime by the time average of \(R_t\) over \(\mathcal T_{\text{mid}}\) (defined in the following paragraph) denoted \(\bar R_{\text{mid}}\).

We next partition the cross-sectional time index \(\{1,\dots,T\}\) into three disjoint regimes based on a single per-date aggregate of the panel of raw daily log returns. The partition is defined on the raw returns directly, so it is computed independently of the univariate stage. The purpose is to define the middle-regime time-averaging window for \(\bar R_{\text{mid}}\) and the convex-combination weights for \(R_{ss}\). Let \(\bar r_t = N^{-1} \sum_{i=1}^N r_{i,t}^{\text{raw}}\) denote the cross-sectional mean log return on date \(t\), and let \(q_{q_{\mathrm{tail}}}(\bar r)\) and \(q_{1-q_{\mathrm{tail}}}(\bar r)\) be the empirical \(q_{\mathrm{tail}}\) and \((1-q_{\mathrm{tail}})\) quantiles of \(\{\bar r_t\}_{t=1}^T\), where the classification threshold \(q_{\mathrm{tail}}\) is a tunable parameter of the analysis (set to \(q_{\mathrm{tail}} = 0.05\) throughout the paper). The three regimes are 
\[
\mathcal T_{\text{upper}} = \{t : \bar r_t > q_{1-q_{\mathrm{tail}}}(\bar r)\},\quad
\mathcal T_{\text{lower}} = \{t : \bar r_t < q_{q_{\mathrm{tail}}}(\bar r)\},\quad
\mathcal T_{\text{mid}}   = \{1,\dots,T\} \setminus (\mathcal T_{\text{upper}} \cup \mathcal T_{\text{lower}}).
\] 
By construction the three regimes partition the time index into \(q_{\mathrm{tail}}\), \(1 - 2 q_{\mathrm{tail}}\), \(q_{\mathrm{tail}}\), so at the \(q_{\mathrm{tail}} = 0.05\) default the middle regime contains exactly \(90\%\) of the trading dates, the upper tail \(5\%\), and the lower tail \(5\%\). The interpretation is market-direction-based: tail dates are the days on which the \emph{cross-section as a whole} moved decisively (up for \(\mathcal T_{\text{upper}}\), down for \(\mathcal T_{\text{lower}}\)), and middle dates are the days on which the cross-sectional average return was within its central \(90\%\) range.  The purpose of this construction is two-fold: (i) the central-regime correlation matrix \(\bar R_{\text{mid}}\) is the time average of the DCC conditional correlation \(R_t\) restricted to \(\mathcal T_{\text{mid}}\), and (ii) the convex combination weights \((w_l, w_m, w_u) = (q_{\mathrm{tail}}, 1 - 2\,q_{\mathrm{tail}}, q_{\mathrm{tail}})\) used to form the steady-state dependence matrix \(R_{ss}\) are derived from the same regime-frequency interpretation.

The multivariate Gaussian likelihood used here is the standard QMLE target with \citet{BollerslevWooldridge1992} sandwich standard errors. The latter are robust to misspecification of the innovation distribution under the QMLE consistency conditions of \citet{White1982}, and are reported in the empirical section as the inferential object of choice. We adopt the standard \texttt{rmgarch} implementation of Engle parameterization here. However, we recommend \citet{Aielli2013} cDCC reparameterization for any engineering or forecasting use of the fitted model. A complementary line of work addresses the estimation for the broader stochastic-volatility family with intractable likelihoods via approximate Bayesian computation rather than
QMLE. For example, \citet{VankovEtAl2019} develop an ABC-based auxiliary particle filter for stochastic-volatility models whose likelihood cannot be evaluated in closed form, providing an inferentially principled alternative for cases where the multivariate aggregation step makes the joint likelihood intractable.

\subsection{Tail Pairwise Dependence Matrix}\label{sec-tpdm}

For the two tail regimes we replace the moment-based DCC estimator with the TPDM of \citet{CooleyThibaud2019}. Working under the assumption that the vector of standardized residuals is \emph{multivariate regularly varying} with tail index \(\alpha>0\), the TPDM replaces the second-moment correlation with a tail moment of a pairwise angular decomposition. Concretely, for each pair \((i,j)\) and each \(t\) one forms the radial and angular components 
\[ 
R_{ij,t}=\sqrt{\,\tilde z_{i,t}^{2}+\tilde z_{j,t}^{2}\,},\qquad W_{ij,t}^{(1)}=\tilde z_{i,t}/R_{ij,t},\qquad W_{ij,t}^{(2)}=\tilde z_{j,t}/R_{ij,t}, 
\] 
where \(\tilde z_{i,t}\) is the result of transforming the empirical cumulative distribution of \(z_{i,t}\) by the inverse Fréchet\((\alpha=2)\) function, so that \(\tilde z_{i,t}\) has tail index \(2\) exactly. The \((i,j)\) entry of the TPDM \(\widehat\Sigma^{\text{TPD}}\) is then 
\[ 
\widehat\Sigma^{\text{TPD}}_{ij}    = \frac{2}{|\mathcal I_{ij}(u)|}\sum_{t\in \mathcal I_{ij}(u)} W^{(1)}_{ij,t}\, W^{(2)}_{ij,t},\qquad    \mathcal I_{ij}(u)=\{t:R_{ij,t}>u\}, 
\]
with \(u\) a high empirical radial threshold (the \(95\)th percentile of pooled radii in our implementation). The factor of \(2\) is the \(\alpha=2\) normalization that places the diagonal at unity and makes the off-diagonal entries lie in \([-1,1]\). Three preprocessing steps are required before the estimator is applied: 1) an empirical CDF smoothing by beta-kernel convolution to avoid the rank ties that arise in small samples; 2) the Fréchet\((\alpha=2)\) probability-integral transform that places \(\tilde z_{i,t}\) on the positive orthant with marginal tail index exactly \(2\); and, 3) following \citet{mhatre2024transformed}, a column-mean-centering of the Fréchet-transformed series followed by clipping any resulting negative entries back to zero. The mean-centering step targets the well-documented positive bias of tail-dependence estimators under weak dependence \citep{HuserEtAl2016}. \citet{mhatre2024transformed} validate the mean-centering through a simulation study in their Supplementary Material that shows the centering materially reduces bias in the estimate. The justification is empirical rather than analytical, and the clipping step keeps the post-centered data on the positive orthant that the transformed-linear framework of \citet{mhatre2024transformed} requires. The upper-tail and lower-tail TPDMs are estimated on the same cross-section after, respectively, retaining the positive transformed data and negating the series before the transform.

We estimate the TPDM on the univariately-filtered residual stream \(\widehat z_{i,t}\) applied to the raw returns under the fitted GARCH(1,1) models. The resulting upper- and lower-tail estimators \(\widehat\Sigma^{\text{TPD},z}_{u}\) and \(\widehat\Sigma^{\text{TPD},z}_{l}\) capture the \emph{total} joint tail dependence in the panel, and are the natural tail-side counterpart of the central-regime \(\bar R_{\text{mid}}\) in the steady-state construction of  \autoref{sec-steady-state-method}. Both tail estimators feed the steady-state convex combination.

\subsubsection{Probabilistic Foundation: Multivariate Regular
Variation}\label{probabilistic-foundation-multivariate-regular-variation}

The TPDM is a sample analogue of an object in extreme value theory rather than a moment of the central distribution, and its interpretation requires the framework of multivariate regular variation (MRV). We assume that the vector of standardized residuals \(z_t\) is multivariate regularly varying with tail index \(\alpha > 0\): there exists a sequence \(a_n \to \infty\) and a non-degenerate Radon measure \(\nu\) on 
\([-\infty, \infty]^N \setminus \{0\}\), called the \emph{exponent measure}, such that for every Borel set \(A\) bounded away from the origin, 
\[ 
n\, P(a_n^{-1}\, z_t \in A) \,\to\, \nu(A). 
\] 
The exponent measure is homogeneous of degree \(-\alpha\), \(\nu(t \cdot A) = t^{-\alpha}\, \nu(A)\), and admits the polar-coordinate decomposition 
\[ 
\nu(\mathrm d r, \mathrm d \omega) \,=\, c\, r^{-\alpha-1}\, \mathrm d r \otimes H(\mathrm d \omega), 
\]
where \(H\) is a finite measure on the unit sphere \(\mathcal S^{N-1}\) called the \emph{angular} (or \emph{spectral}) measure. The angular measure encodes the dependence structure of the joint extremes. Specifically, its support determines which combinations of components can be simultaneously large, and its mass distribution determines how strongly they covary in the tails. Under this framework, the off-diagonal \((i,j)\) entry of the TPDM is the bivariate angular moment 
\[ 
\Sigma^{\text{TPD}}_{ij} \,=\, \int_{\mathcal S^{N-1}} \omega_i\, \omega_j\, H(\mathrm d \omega) 
\]
when the radial component is normalized so that \(\alpha = 2\). This is the construction of \citet{CooleyThibaud2019}, and the underlying MRV machinery is developed in \citet{Resnick2007} and \citet{EmbrechtsEtAl1997}. The ECDF-smoothing and inverse-Fréchet steps in our algorithmic definition of \(\tilde z_{i,t}\) above are precisely the device that forces the marginals to be exactly Pareto with \(\alpha = 2\), so that the angular-moment interpretation is well-defined regardless of the marginal tail index of the original \(z_{i,t}\). The Mhatre-Cooley bias-correction step described above is a location shift on the Frèchet-transformed series and leaves the angular measure \(H\) at infinity, and therefore the population target \(\Sigma^{\text{TPD}}_{ij}\) of the angular moment, unchanged.

Two consequences for interpretation follow. First, \(\Sigma^{\text{TPD}}_{ij}\) is a parameter of the joint extremal distribution rather than a statistic of the central regime, so values near zero can be read as evidence of asymptotic tail independence even when the central correlation is large. Second, the TPDM is determined by the angular measure \(H\) alone and is invariant under monotone marginal transformations. The radial information is integrated out by the threshold and averaging step in the estimator. This gives the TPDM a copula-like flavor without the parametric commitment of a copula family. 

\subsubsection{Asymptotic Properties}\label{sec-tpdm-asymptotic}

Under the MRV assumption together with mild regularity on the angular measure and on the rate at which the radial threshold is raised, Cooley and Thibaud (2019, Theorem 4) establish consistency of the radial-threshold estimator \(\widehat\Sigma^{\text{TPD}}_{ij}\) as the threshold \(u \to \infty\) and the number of exceedances \(|\mathcal I_{ij}(u)| \to \infty\) at appropriate relative rates. Under the same conditions, the off-diagonal entries satisfy a central-limit-type result with rate of convergence \[ \widehat\Sigma^{\text{TPD}}_{ij} - \Sigma^{\text{TPD}}_{ij} \,=\, O_p\!\left(|\mathcal I_{ij}(u)|^{-1/2}\right), \] with a leading bias term of order \(u^{-\alpha}\) that decreases as the threshold is raised. The two scales operate in tension, where a higher threshold reduces bias by isolating the tail more cleanly but reduces the number of exceedances and so inflates the variance. For our cross-section of  3,228 observed time points, the \(95\)th-percentile radial threshold yields roughly \(|\mathcal I_{ij}(u)| \approx 161\) exceedances per pair, giving a per-entry standard error of order \(1/\sqrt{161} \approx 0.07\). We use two times this scale below as a soft significance threshold for which off-diagonal differences across regimes are statistically meaningful and which are within the margin of sampling noise. The \(0.14\) figure is back-of-envelope as it ignores the dependence of the entry-level variance on the angular measure \(H\) and treats every pair symmetrically, while the actual variance of \(\widehat\Sigma^{\text{TPD}}_{ij}\) scales with the second angular moment of \((W^{(1)}_{ij}, W^{(2)}_{ij})\) on the radial-exceedance set. 

\subsection{Steady-State Dependence by Convex Combination}\label{sec-steady-state-method}
While the three regime-specific dependence matrices are informative in their own right, many applications (unconditional portfolio optimization, long-horizon risk budgeting, benchmark comparisons) require a single correlation matrix that summarizes long-run dependence. We obtain such an estimator by convex combination of the three regime matrices: 
\[
R_{ss} = w_l\, \widehat\Sigma^{\text{TPD}}_{\text{lower}} + w_m\, \bar R_{\text{mid}} + w_u\, \widehat\Sigma^{\text{TPD}}_{\text{upper}},
\qquad w_l + w_m + w_u = 1,\; w_\bullet \ge 0.
\] 
The weights admit a natural interpretation as the unconditional probabilities of being in each regime. Under the \(q_{\mathrm{tail}} = 0.05\) marginal threshold adopted above, \(w_l = w_u = q_{\mathrm{tail}} = 0.05\) and \(w_m = 1 - 2\, q_{\mathrm{tail}} = 0.90\) is the symmetric baseline. Using the empirical regime frequencies, or weights derived from a regime-switching Markov chain fit to the classification indicator, are natural extensions that leave the decomposition structure unchanged. By construction, the convex combination \(R_{ss}\) is a symmetric matrix with unit entries along the main diagonal. 

We are deliberately not claiming that \(R_{ss}\) is the marginal expectation under a single coherent probabilistic model. The tail-regime components \(\widehat\Sigma^{\text{TPD}}_{l}\) and \(\widehat\Sigma^{\text{TPD}}_{u}\) are angular moments of the spectral measure under multivariate regular variation and they live in a different probability framework from \(\bar R_{\text{mid}}\) and are not conditional expectations of the same population \(R_t\). The three components share the algebraic structure of correlation matrices, in other words they are symmetric have a diagonal of one's and off-diagonals in \([-1, 1]\) under the \(\alpha = 2\) normalization. These features make the convex combination algebraically well-defined and produces a valid dependence matrix as output, but they do not share a single common population target. The right reading of \(R_{ss}\) is as an aggregate dependence summary that injects tail-dependence information from the two tail TPDMs into a central-regime correlation summary, with the weights interpretable as regime frequencies.

A natural forecasting extension under the heuristic reading treats \(\{S_t\}\) as a three-state Markov chain in the spirit of \citet{Hamilton1989}, with transition matrix \(P = (P_{kj})\) and stationary distribution \(\pi\). The single-period-ahead dependence aggregator is then 
\[
\sum_{j \in \{l, m, u\}} P_{kj}\, R_j,
\] 
which, together with an estimate of the transition matrix from the empirical regime sequence, supplies a forward-looking analogue of the steady-state matrix. The closely related threshold-process viewpoint \citep{Tong1990} makes the regime indicator deterministically dependent on the lagged residual vector, which is consistent with the rule we use here. As with the steady-state version, the Markov-update aggregator is best understood as a heuristic combination of regime-specific dependence summaries rather than as a strict forecast under a single dependence model.

\subsection{Assessing the Steady-State Estimator, the Three Regimes, and Tail Dependence}\label{assess}

Visual heat maps are presented for the lower TPDM, the average DCC matrix, and the upper TPDM. Correlation network graphs are also presented for each regime. Further, heat maps are used to visualize the differences in the TPDM structure between the upper and lower regimes, specifically the contrast $\widehat\Sigma^{\text{TPD}}_u - \widehat\Sigma^{\text{TPD}}_l$ is plotted. 

The Spearman rank based estimated correlation matrix, $R^{s}$, is computed from the raw daily log returns and serves as benchmarks against which the regime-aware steady-state estimator, \(R_{ss}\), is compared. The pairwise differences in the estimated correlations are visually compared with a smoothed density estimate broken out by the ETF. 

As a formal test of tail dependence, we turn to the Gaussian copula correlation matrix as our base method. The Gaussian copula is the canonical industry baseline for joint dependence in equity returns \citep{McNeilEtAl2015}, but it cannot capture the dependence in the tails of the distribution. We rely on the independence in the tails of the Gaussian copula model to test the significance of the tail dependence in our upper and lower regimes. The Gaussian-copula correlation matrix is fit to the standardized residual stream \(\widehat z_{i,t}\) that feeds the TPDM. For each equity \(i\) we transform \(\widehat z_{i,t}\) first to uniform marginals via its empirical CDF and then to standard-normal marginals via the inverse normal CDF, \(W_{i,t} = \Phi^{-1}(F_i(\widehat z_{i,t}))\). The Gaussian-copula correlation matrix \(\widehat R^{\text{gp}}\) is then the Pearson correlation of \(W\). 

Formal simulation-based inference is conducted to compare the TPDM-based dependence estimators, $\widehat\Sigma^{\text{TPD}}_l$ and $\widehat\Sigma^{\text{TPD}}_u$, with the Gaussian copula–based estimator, $R^{GP}$, each fit to the standardized residuals as noted above. Under the Gaussian copula null hypothesis, tail independence implies that each TPDM is, in expectation, estimating the identity matrix. The testing algorithm is as follows:
\begin{itemize}
    \item Simulate \(M\) panels of length
\(T\) from a Gaussian copula with correlation
\(\widehat R^{\text{gp}}\)
    \item Follow the process to compute $\widehat\Sigma^{\text{TPD}}_u$ and $\widehat\Sigma^{\text{TPD}}_l$, for each of the \(M\) panels.
    \item Accumulate the per-entry simulated distribution of TPDM values under the Gaussian-copula null hypothesis
    \item Off-diagonal entries of the observed TPDM that exceed the \(95\)th percentile of their MC simulated null distribution are flagged as rejecting the Gaussian copula at the \(5\%\) level, separately for the upper and lower tails.
    \item Benhamini-Hocberg False Discover Rate (BH-FDR) methods are implemented to understand our findings across the 528 univariate tests for each tail. 
\end{itemize}
We run the test at two radial thresholds, \(u = 0.95\) and \(u = 0.975\) so we can ascertain the threshold-sensitivity of the rejection counts.

The tail-independence property of the Gaussian copula is an asymptotic statement that holds only in the limit \(u \to 1\). Accordingly, both the sign and magnitude of any threshold-induced dependence observed in the rejection counts constitute the principal source of inferential information. Persistently elevated rejection counts as the threshold is made more stringent imply that the asymptotic discrepancy between the Gaussian-copula-based prediction and the empirical joint tail dependence is non-negligible. Conversely, changes in rejection counts as \(u\) increases encode information about the trade-off between (i) reduced statistical power, arising from the decreasing number of per-pair exceedances at higher \(u\), and (ii) stronger asymptotic discrimination as \(u\) approaches its upper limit. 

The simulation employs \(M = 5000\) Monte Carlo replicates, yielding an approximate \(\pm 1.9\) percentage-point margin of error for the per-pair 95th-percentile cutoff corresponding to a \(\pm 1.96\) Wald confidence band around the cutoff proportion. The qualitative conclusions based on the total number of rejections are stable for substantially smaller \(M\), and increasing \(M\) to 5000 further refines the per-pair quantile estimates, at the expense of increased initial computational time. Further we apply the Benjamini-Hochberg False Discovery Rate (BH-FDR) adjustment for multiple comparisons. 

\section{Data and Software}\label{data}

Our empirical analysis focuses on the constituents of two commodity-benchmark ETFs: the Invesco Water Resources ETF (CGW) and the Energy Select Sector SPDR (XLE). This yields a panel that is both economically meaningful and spans multiple sectors with distinctly different sensitivities to macroeconomic shocks. We set out the equity cross-section under study in detail, describe the preprocessing and data-cleaning steps applied to the raw price series, explain the deterministic ticker-ordering convention that is used consistently across all figures and tables.

The cross-sectional sample comprises the constituents of the Invesco Water Resources ETF (CGW) and the Energy Select Sector SPDR (XLE). We exclude fourteen equities due to a high proportion of missing observations, resulting in a final cross-section of thirty-three securities. In addition, we account for a ticker change that occurred in 2025 by substituting the new ticker for the original and reindexing the corresponding time series across the entire sample period.

A canonical ticker ordering is generated in a deterministic manner from the published ETF holdings files by parsing the per-security weight strings. This procedure ensures that all TICKER-axis visualizations and coefficient panels follow a consistent arrangement, in which the energy-sector block precedes the water-ETF block. 

The raw log daily return of each asset is computed and denoted by \(r_{i,t}^{\text{raw}}\) for \(i \in \{1,\dots,N\}\) at time \(t \in \{1,\dots,T\}\). The winsorized preprocessing step results in winsorized returns denoted by \(r_{i,t}^{\text{wins}}\). Winsorization is based on each returns empirical \(1\%\) and \(99\%\) quantiles. These values are listed in Table \ref{tab:garch_coefficients}.  

The 3,587 days in our sample are partitioned into a middle-90\% subsample of 3,228 days by applying the cross-sectional-mean rule, leaving 179 days in each tail. Recall that a date \(t\) is classified according to the position of its cross-sectional mean return \(\bar r_t = N^{-1} \sum_i r_{i,t}\) within the empirical distribution of \(\{\bar r_t\}_t\).

The empirical implementation is conducted in \textsf{R} (version 4.3.1). The primary packages employed are \texttt{rugarch} \citep{GhalanosRugarch} for the estimation of univariate GARCH models, and \texttt{rmgarch} \citep{GhalanosRmgarch} for the estimation of multivariate GARCH specifications and the Dynamic Conditional Correlation (DCC) model. Financial time series are obtained and pre-processed using \texttt{quantmod}, whereas data manipulation and visualization are performed with \texttt{ggplot2} and the broader \texttt{tidyverse} ecosystem. In addition, \texttt{ggraph} is utilized to produce graphical representations that facilitate visual comparison of the estimated dependence structures. All source code and detailed implementation procedures are publicly accessible via the CoFES GitHub repository.

\section{Results}\label{results}
With the preprocessing procedures and data characterization now complete, we proceed to the analytical phase as outlined in Figure \ref{fig-pipeline}.

\subsection{Univariate GARCH Fits}\label{univariate-garch-fits}

\begin{figure}[H]

\centering{

\includegraphics[width=1\linewidth,height=\textheight,keepaspectratio]{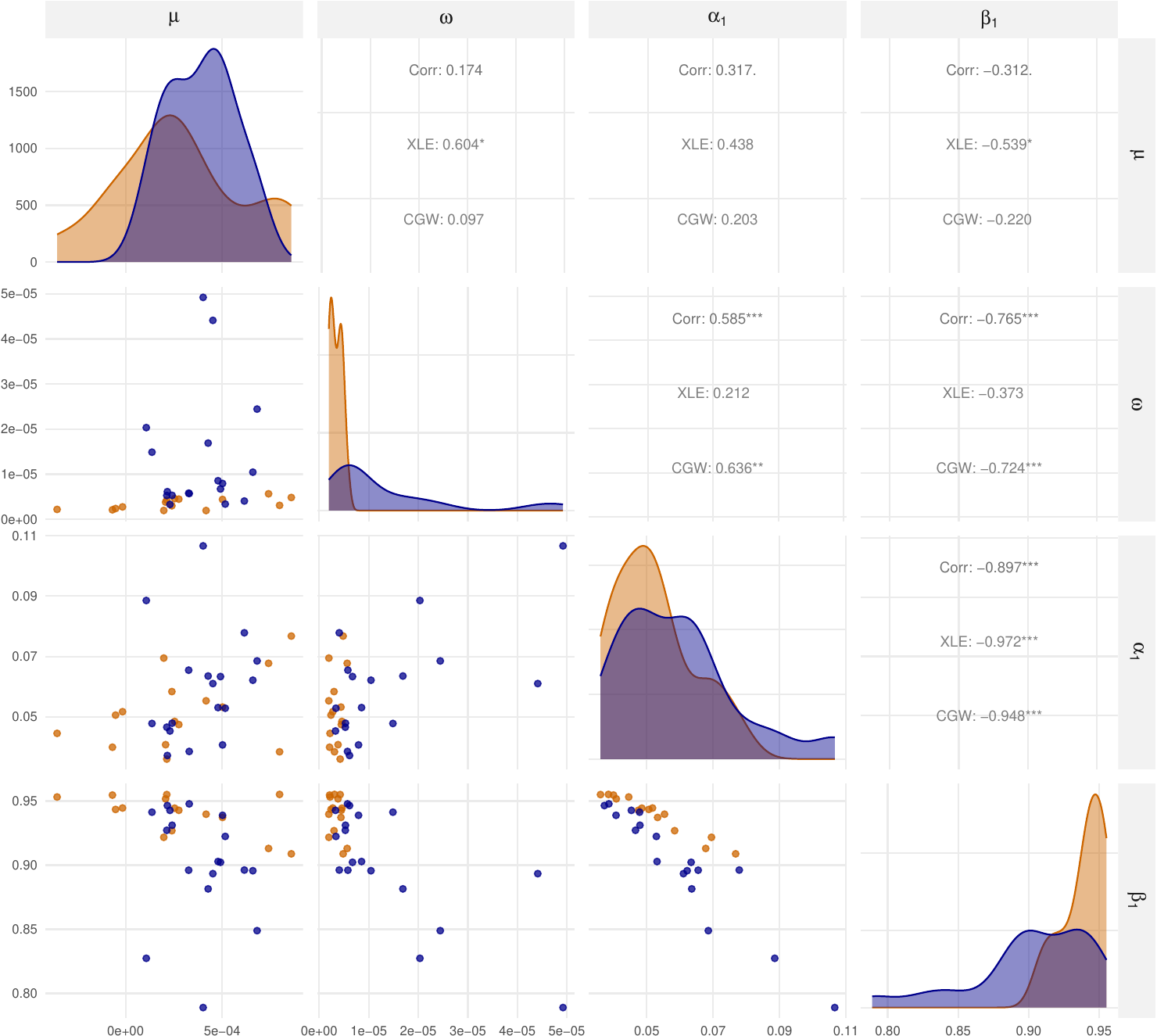}

}

\caption{\label{fig-garch-coef-pairs}Pairwise scatterplots of the estimated GARCH(1,1) coefficients \(\mu_i\), \(\omega_i\), \(\alpha_{i,1}\), and \(\beta_{i,1}\) across the 33-stock cross-section. Points are colored by ETF membership using the manuscript-wide palette (XLE dark orange, CGW dark blue). Lower-triangle panels show the per-pair scatterplots, upper-triangle panels show the Pearson correlation coefficient broken out by ETF and overall, and diagonal panels show the marginal density of each coefficient by ETF. Axis tick labels and panel strip labels are rendered in publication-quality typography with the coefficient names as Greek letters with subscripts.}

\end{figure}

We estimate a conventional GARCH(1,1) model with a constant conditional mean and Gaussian innovations for each security using quasi–maximum likelihood on the winsorized return series. The resulting parameter estimates are reported in Table \ref{tab:garch_coefficients} and depicted in Figure \ref{fig-garch-coef-pairs}.

The conditional volatilities \(\widehat\sigma_{i,t}\) and standardized residuals \(\widehat z_{i,t}\) for each asset are subsequently obtained by filtering the original raw return series using the parametric models estimated on the winsorized return series. In parallel, the winsorized standardized residuals \(\widehat z_{i,t}^{\,\text{wins}}\) are derived from the estimated winsorized conditional volatilities \(\widehat\sigma_{i,t}^{\text{wins}}\), which are computed by filtering the winsorized return series. 

For clarity, the Dynamic Conditional Correlation (DCC) model is estimated using the winsorized standardized residuals, while the Tail Pairwise Dependence Matrix (TPDM) methodology is applied to the (non-winsorized) standardized residuals. Moreover, the benchmark correlation measures, Pearson and Spearman, or computed from the original data and the Gaussian-copula-based correlations are constructed from standardized residuals to directly compare with the TPDM structure. 

\begin{longtable}[]{@{}
  >{\raggedright\arraybackslash}p{(\linewidth - 14\tabcolsep) * \real{0.0707}}
  >{\raggedleft\arraybackslash}p{(\linewidth - 14\tabcolsep) * \real{0.0909}}
  >{\raggedleft\arraybackslash}p{(\linewidth - 14\tabcolsep) * \real{0.0909}}
  >{\raggedleft\arraybackslash}p{(\linewidth - 14\tabcolsep) * \real{0.15}}
  >{\raggedleft\arraybackslash}p{(\linewidth - 14\tabcolsep) * \real{0.08}}
  >{\raggedleft\arraybackslash}p{(\linewidth - 14\tabcolsep) * \real{0.15}}
  >{\raggedleft\arraybackslash}p{(\linewidth - 14\tabcolsep) * \real{0.08}}
  >{\raggedleft\arraybackslash}p{(\linewidth - 14\tabcolsep) * \real{0.15}}@{}}
\caption{Estimated GARCH(1,1) coefficients by TICKER (Gaussian innovations, fit on the winsorized series). \(q_{lo}\) and \(q_{hi}\) are the per-series 1\% and 99\% return quantiles used as the winsorization thresholds. \(\widehat\sigma_\infty = \sqrt{\omega / (1 - \alpha_1 - \beta_1)}\) is the implied unconditional standard deviation of the conditional volatility under stationarity. SE(\(\alpha_1\)) and SE(\(\beta_1\)) are Bollerslev-Wooldridge robust sandwich standard errors. Rows are sorted in \texttt{ticker\_order}.\label{tab:garch_coefficients}}\tabularnewline
\toprule\noalign{}
\begin{minipage}[b]{\linewidth}\raggedright
TICKER
\end{minipage} & \begin{minipage}[b]{\linewidth}\raggedleft
\(q_{lo}\)
\end{minipage} & \begin{minipage}[b]{\linewidth}\raggedleft
\(q_{hi}\)
\end{minipage} & \begin{minipage}[b]{\linewidth}\raggedleft
\(\alpha_1\)
\end{minipage} & \begin{minipage}[b]{\linewidth}\raggedleft
SE(\(\alpha_1\))
\end{minipage} & \begin{minipage}[b]{\linewidth}\raggedleft
\(\beta_1\)
\end{minipage} & \begin{minipage}[b]{\linewidth}\raggedleft
SE(\(\beta_1\))
\end{minipage} & \begin{minipage}[b]{\linewidth}\raggedleft
\(\widehat\sigma_\infty\)
\end{minipage} \\
\midrule\noalign{}
\endfirsthead
\toprule\noalign{}
\begin{minipage}[b]{\linewidth}\raggedright
TICKER
\end{minipage} & \begin{minipage}[b]{\linewidth}\raggedleft
\(q_{lo}\)
\end{minipage} & \begin{minipage}[b]{\linewidth}\raggedleft
\(q_{hi}\)
\end{minipage} & \begin{minipage}[b]{\linewidth}\raggedleft
\(\alpha_1\)
\end{minipage} & \begin{minipage}[b]{\linewidth}\raggedleft
SE(\(\alpha_1\))
\end{minipage} & \begin{minipage}[b]{\linewidth}\raggedleft
\(\beta_1\)
\end{minipage} & \begin{minipage}[b]{\linewidth}\raggedleft
SE(\(\beta_1\))
\end{minipage} & \begin{minipage}[b]{\linewidth}\raggedleft
\(\widehat\sigma_\infty\)
\end{minipage} \\
\midrule\noalign{}
\endhead
\bottomrule\noalign{}
\endlastfoot
XOM & -0.0452 & 0.0431 & 0.0695 & 0.7632 & 0.9217 & 0.8389 & 0.0149 \\
CVX & -0.0439 & 0.0416 & 0.0584 & 0.0298 & 0.9269 & 0.0448 & 0.0143 \\
COP & -0.0572 & 0.0547 & 0.0554 & 0.1070 & 0.9397 & 0.1152 & 0.0199 \\
WMB & -0.0602 & 0.0594 & 0.0678 & 0.2383 & 0.9130 & 0.4975 & 0.0172 \\
EOG & -0.0630 & 0.0645 & 0.0533 & 0.0302 & 0.9372 & 0.0422 & 0.0215 \\
VLO & -0.0662 & 0.0607 & 0.0385 & 0.0031 & 0.9552 & 0.0043 & 0.0220 \\
SLB & -0.0610 & 0.0607 & 0.0400 & 0.0055 & 0.9547 & 0.0073 & 0.0198 \\
BKR & -0.0623 & 0.0618 & 0.0475 & 0.0187 & 0.9428 & 0.0292 & 0.0214 \\
OKE & -0.0621 & 0.0579 & 0.0768 & 0.0183 & 0.9089 & 0.0367 & 0.0183 \\
EQT & -0.0735 & 0.0775 & 0.0485 & 0.0222 & 0.9445 & 0.0303 & 0.0256 \\
OXY & -0.0668 & 0.0718 & 0.0506 & 0.0978 & 0.9435 & 0.1105 & 0.0201 \\
HAL & -0.0697 & 0.0676 & 0.0408 & 0.0077 & 0.9517 & 0.0110 & 0.0226 \\
DVN & -0.0759 & 0.0779 & 0.0517 & 0.0558 & 0.9446 & 0.0621 & 0.0273 \\
CTRA & -0.0657 & 0.0672 & 0.0361 & 0.0033 & 0.9551 & 0.0030 & 0.0219 \\
APA & -0.0812 & 0.0835 & 0.0446 & 0.0087 & 0.9532 & 0.0102 & 0.0311 \\
AWK & -0.0366 & 0.0309 & 0.0529 & 0.0618 & 0.9224 & 0.0894 & 0.0117 \\
WTRG & -0.0382 & 0.0332 & 0.0655 & 0.0032 & 0.8961 & 0.0057 & 0.0123 \\
ECL & -0.0411 & 0.0382 & 0.0779 & 0.0344 & 0.8962 & 0.1274 & 0.0125 \\
WTS & -0.0460 & 0.0511 & 0.0622 & 0.0031 & 0.8957 & 0.0061 & 0.0157 \\
AWR & -0.0404 & 0.0381 & 0.0531 & 0.0015 & 0.9029 & 0.0038 & 0.0139 \\
BMI & -0.0541 & 0.0572 & 0.0685 & 0.0263 & 0.8490 & 0.0712 & 0.0172 \\
CWT & -0.0417 & 0.0390 & 0.0466 & 0.0038 & 0.9272 & 0.0044 & 0.0142 \\
VMI & -0.0469 & 0.0487 & 0.0479 & 0.0057 & 0.9311 & 0.0078 & 0.0159 \\
MWA & -0.0669 & 0.0631 & 0.0408 & 0.0025 & 0.9389 & 0.0053 & 0.0198 \\
FELE & -0.0522 & 0.0521 & 0.0635 & 0.0079 & 0.8815 & 0.0160 & 0.0175 \\
SJW & -0.0475 & 0.0434 & 0.0885 & 0.0188 & 0.8274 & 0.0437 & 0.0156 \\
OTTR & -0.0417 & 0.0400 & 0.0634 & 0.0029 & 0.9023 & 0.0055 & 0.0140 \\
MSEX & -0.0452 & 0.0464 & 0.0454 & 0.7171 & 0.9428 & 0.9628 & 0.0166 \\
HWKN & -0.0603 & 0.0645 & 0.0386 & 0.0057 & 0.9478 & 0.0082 & 0.0205 \\
LNN & -0.0609 & 0.0616 & 0.0372 & 0.0041 & 0.9465 & 0.0054 & 0.0194 \\
CWCO & -0.0632 & 0.0660 & 0.1066 & 0.0237 & 0.7890 & 0.0555 & 0.0217 \\
TTI & -0.1055 & 0.1195 & 0.0478 & 0.0250 & 0.9413 & 0.0083 & 0.0371 \\
ERII & -0.0878 & 0.0986 & 0.0611 & 0.0648 & 0.8934 & 0.1585 & 0.0311 \\
\end{longtable}

\subsection{Middle Region and Dynamic Conditional Correlation, \(\bar R_{\text{mid}}\)}\label{middle-region-and-dynamic-conditional-correlation}

We estimate the DCC model on the winsorized standardized-residual matrix, yielding values of \(\widehat{a}=0.0031\) and \(\widehat{b}=0.9909\), with standard errors of 0.0002 and 0.0009, respectively. The implied persistence is \(\widehat{a}=0.0031 + \widehat{b}=0.9909 = 0.9940\). The persistence sits very close to the unit-root boundary of 1, the well-documented near-integrated regime of equity correlation dynamics. Note, we also fit the asymmetric DCC of \citet{CappielloEtAl2006} with a multivariate Student-t likelihood as a robustness check on this cross-section and found a similarly near-integrated persistence. Hence, the near-unit-root property is robust to the choice of DCC variant within this family rather than a model-selection artifact. The near-integration means that the results of the two-step QMLE consistency degrade and that the standard errors on the parameter estimates should be interpreted with corresponding caution. The principled methodological response is the corrected DCC (cDCC) reparameterization of \citet{Aielli2013}, which restores two-step QMLE consistency under standard regularity conditions. However, the \texttt{rmgarch} implementation we use here does not provide the cDCC implementation, so we flag it as the appropriate principled robustness check. We do note, that the downstream object \(\bar R_{\text{mid}}\), in other words, the time average of the conditional correlation path \(R_t\) over middle-regime dates, depends on the path of \(R_t\) rather than on parameter inference about \((a, b)\), and is therefore robust to the near-integration concern.

Recall that the cross-sectional-mean-based partition \(\mathcal T_{\text{mid}}\) is not employed to constrain the DCC parameter estimation; rather, it is used exclusively to specify the time-averaging window for the central-regime correlation matrix \(\bar R_{\text{mid}}\). Following estimation, we extract \(R_t\), the conditional correlation matrix of the winsorized residuals, for each date in the sample and compute its average over \(\mathcal T_{\text{mid}}\) to obtain \(\bar R_{\text{mid}}\). Thus, in addition to describing the temporal evolution of the typical correlation structure, the DCC model is used to characterize the average correlation in the central regime under the robust specification.

\begin{figure}[H]
\centering{
\pandocbounded{\includegraphics[keepaspectratio]{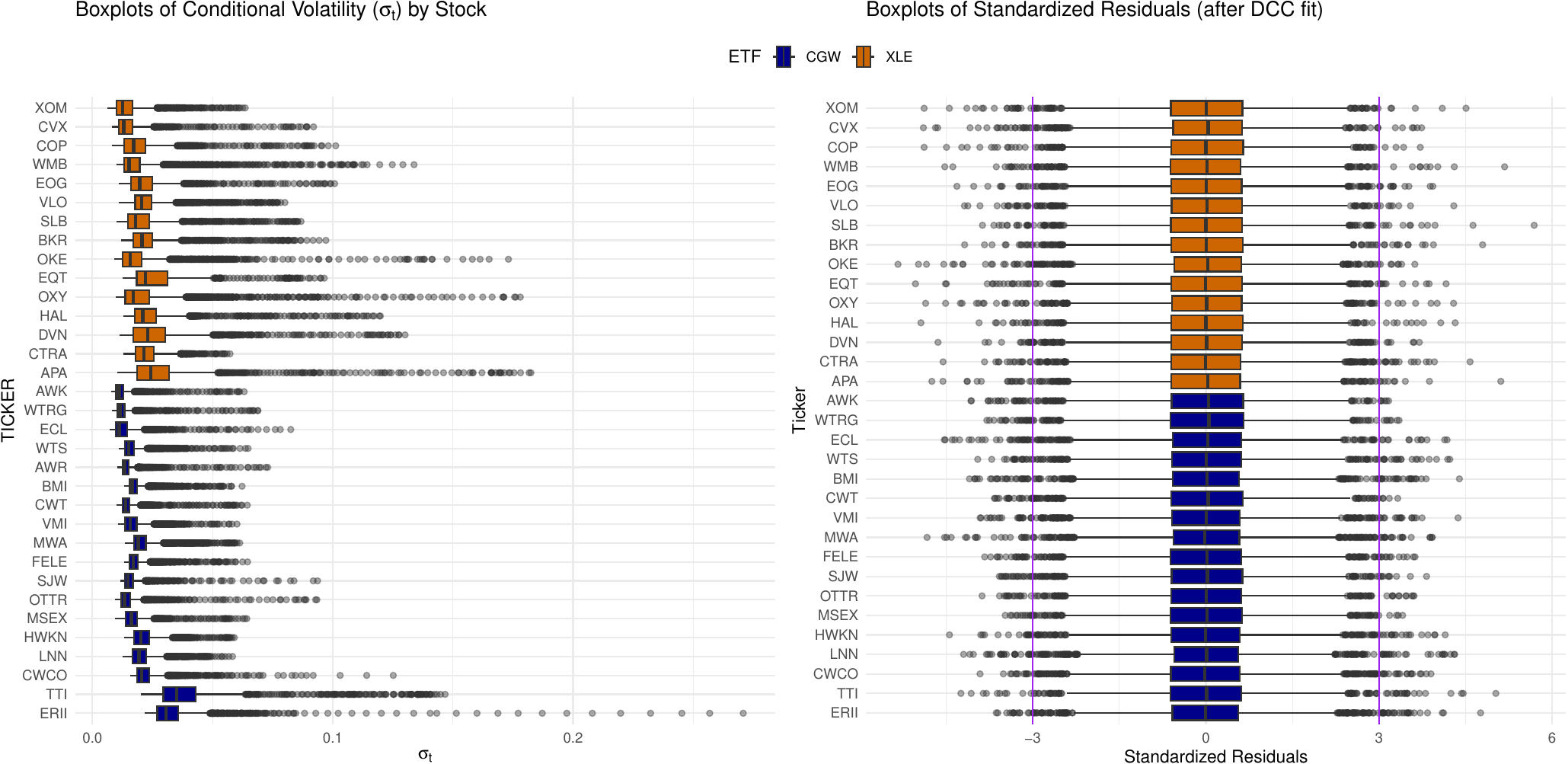}}
}
\caption{\label{fig-vol-and-dccresid-combined}Boxplot of per-ticker conditional volatility from the univariate GARCH stage (left) and boxplot of per-ticker standardized residuals from the full-sample DCC fit (right).}
\end{figure}

Figure~\ref{fig-vol-and-dccresid-combined} displays two descriptive boxplots. The left panel shows the cross-sectional distribution of per-ticker conditional volatilities obtained from the univariate estimation stage, while the right panel presents the per-ticker standardized residuals derived from the full-sample DCC model estimation. The first panel highlights the extent of cross-sectional heterogeneity in conditional volatility levels, whereas the second panel depicts the standardized residuals after applying the DCC filter. 

The DCC model captures the dynamic marginal correlation across the equities. To illustrate how much this association can change over time, a heatmap with 25 evenly-spaced dates sampled from the full DCC fit is provided. In this heatmap we clearly see the dramatic change in market structure from mid to late 2011. In addition, we select the top two equities from each ETF, namely XOM and AWK, and examine the dynamic correlation over our study period. The conditional correlation between XOM and AWK on average is 0.18 and ranges from 0.041 to 0.316. The above average correlations are observed before mid-2016, whereas after that date the correlations are generally below average. 

\subsubsection{Conditional Correlation Distance Over Time}\label{conditional-correlation-distance-over-time}
\begin{figure}[H]
\centering{
\pandocbounded{\includegraphics[keepaspectratio]{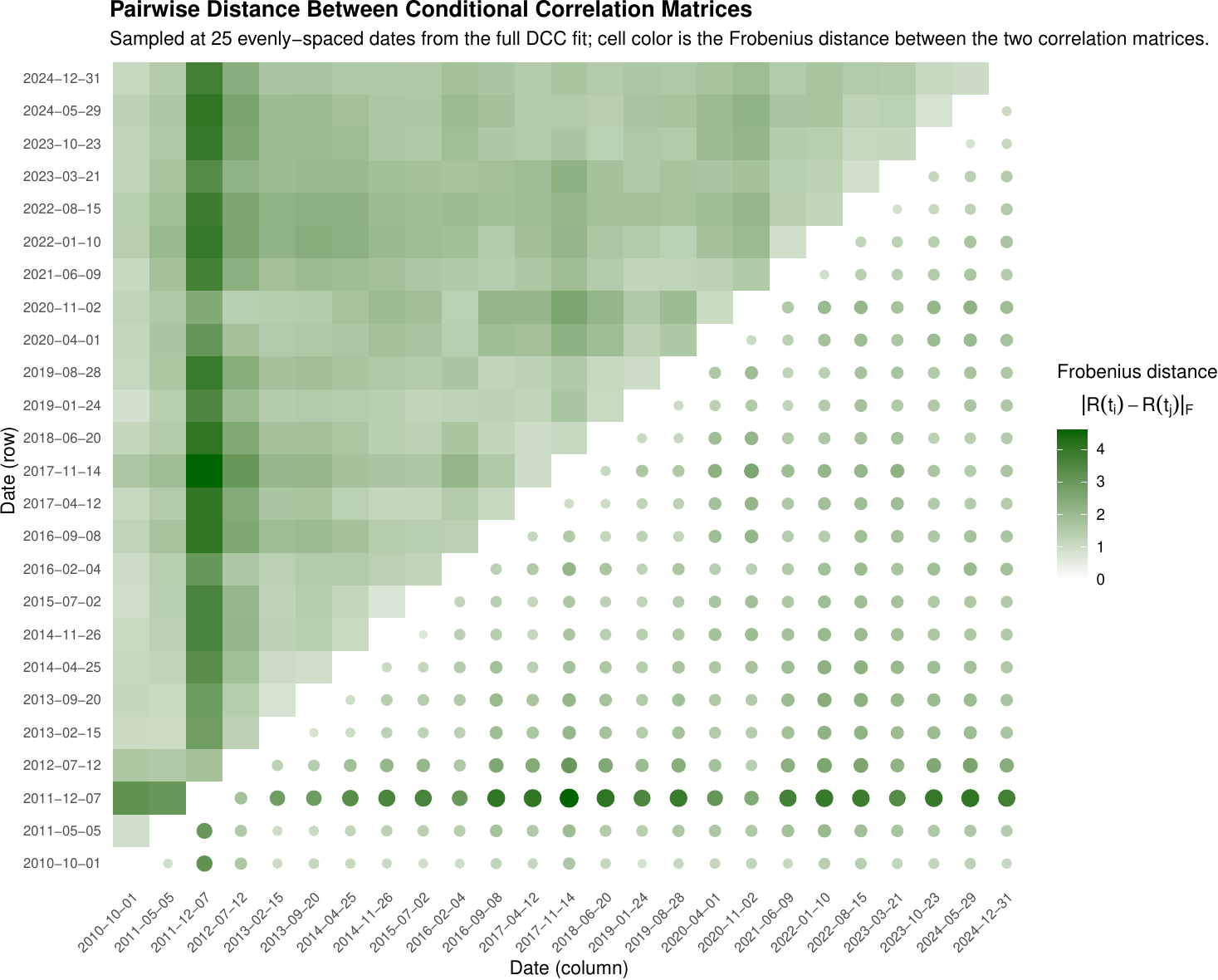}}
}
\caption{\label{fig-dcc-cordist}Pairwise Frobenius distance between the DCC conditional correlation matrices at 25 evenly-spaced dates sampled from the full DCC fit. Cell \((i, j)\) shows \(\lVert R_{t_i} - R_{t_j} \rVert_F\). The matrix is symmetric, so two encodings are used without duplicating information: the visually-upper triangle (and the zero diagonal) is rendered as a colored tile, and the visually-lower triangle is rendered as a colored circle whose color and radius are both tied to the distance on the same white-to-darkgreen scale. Darker / larger cells in either triangle indicate larger displacements in correlation structure between the two dates and the diagonal vanishes by construction. The diverging purple/white/darkgreen scheme used elsewhere in the manuscript for correlations on \([-1, 1]\) is reduced here to its non-negative half because Frobenius distances are by construction \(\ge 0\).}

\end{figure}%

\begin{figure}[H]
\centering{
\pandocbounded{\includegraphics[keepaspectratio]{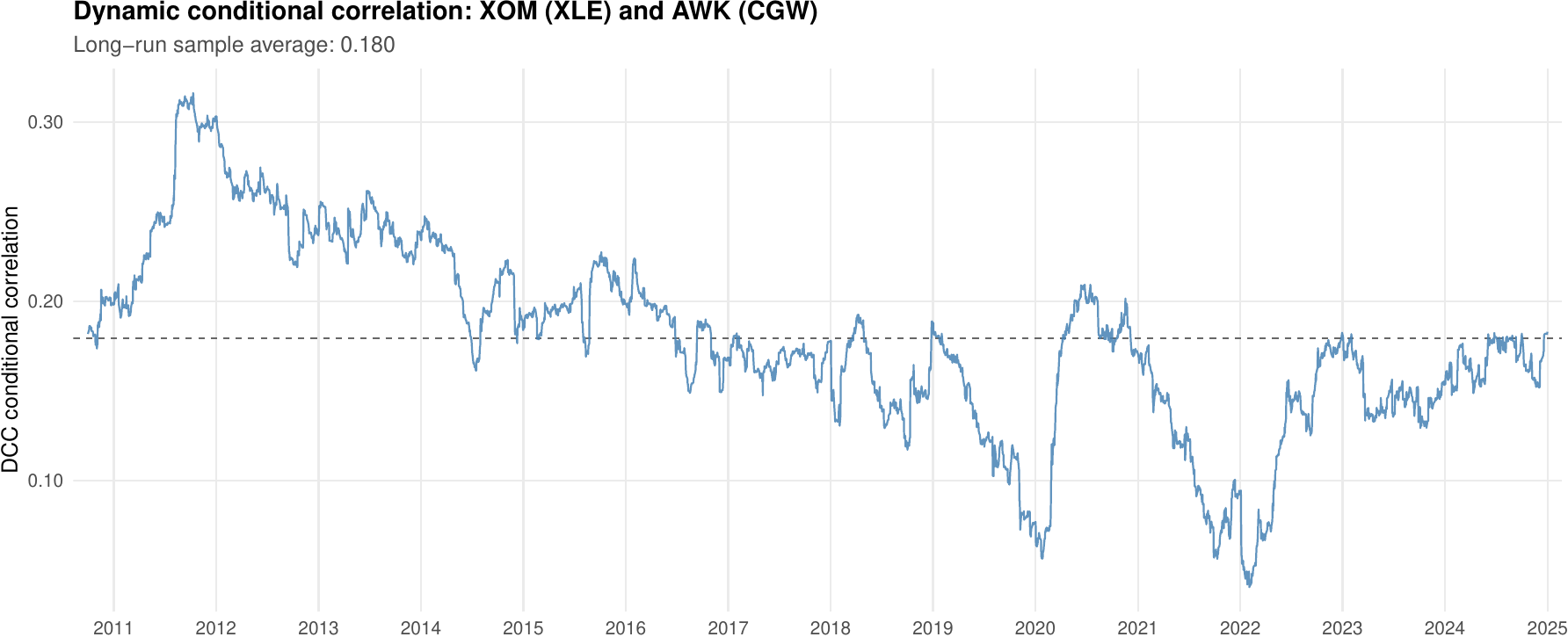}}
}
\caption{\label{fig-dcc-pair-timeseries}DCC conditional correlation between XOM (top-weight XLE name present in the panel) and AWK (top-weight CGW name present in the panel) over the full DCC fit window. The conditional correlation between XOM and AWK on average is 0.18 (dashed horizontal line) and ranges from 0.041 to 0.316.}
\end{figure}

Figure~\ref{fig-dcc-pair-timeseries} plots the DCC conditional correlation between the top-weight XLE equity and the top-weight CGW equity over the entire fit window, as a per-pair complement to the aggregate Frobenius-distance heatmap of Figure~\ref{fig-dcc-cordist}.

\subsection{Tail Pairwise Dependence Matrices}\label{tail-pairwise-dependence-matrices}

\begin{figure}[H]
\centering{
\pandocbounded{\includegraphics[keepaspectratio]{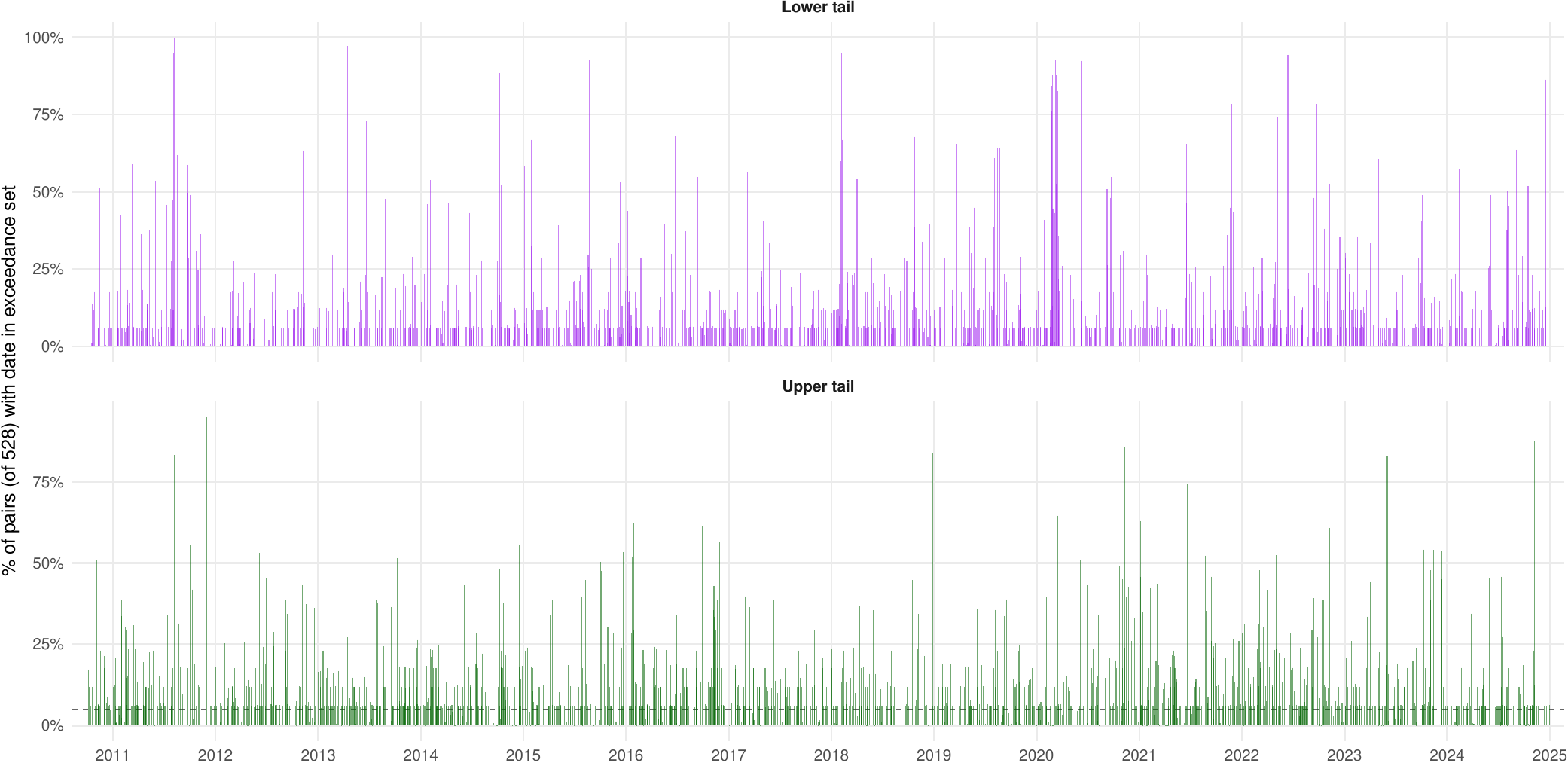}}
}
\caption{\label{fig-tpdm-pair-contribution}Per-date percentage of unordered off-diagonal pairs whose pair-specific top-5\% radial exceedance set includes that date, computed separately for the lower-tail TPDM (top, purple) and the upper-tail TPDM (bottom, dark green). For each pair \((i, j)\) with \(i < j\) we form \(r_{ij,t} = \sqrt{X_{t,i}^2 + X_{t,j}^2}\) from the Fréchet-margin-transformed standardized residuals \(X\) and select dates where \(r_{ij,t}\) exceeds its own 95\% sample quantile; the per-date value is the count of pairs (out of \(N(N-1)/2 = 528\) with \(N = 33\)) for which date \(t\) is in the exceedance set, expressed as a percentage of the total pair count. The horizontal reference line at the independence-baseline expected percentage \(100 \cdot (1 - u) = 5\%\) marks the level a typical day would contribute if exceedances were spread uniformly across dates.}
\end{figure}

We now apply the TPDM construction. The estimation procedure is implemented twice. In the first iteration, the original residuals are used to estimate the upper-tail TPDM, whereas in the second iteration, sign-inverted residuals are employed to estimate the lower-tail TPDM. Since the TPDM does not row-filter the input, each off-diagonal entry \(\widehat\Sigma^{\text{TPD}}_{ij}\) is computed on its own pair-specific top-\((1 - u)\) radial exceedance set, so different pairs select different dates.  To make the time profile of the TPDM's effective sample visible the percentage of the 528 pairs contributing to the calculaton at each time point is plotted in Figure \ref{fig-tpdm-pair-contribution}. If there was no tail-wise dependence present one would expect on average approximately 26/528 or 5\% pairs per date. 

The steady-state estimate of the unconditional correlation structure, \(R_{ss}\), follows immediately from convex combination of the three regime-specific matrices. 

\begin{figure}[H]

\centering{

\includegraphics[width=1\linewidth,height=\textheight,keepaspectratio]{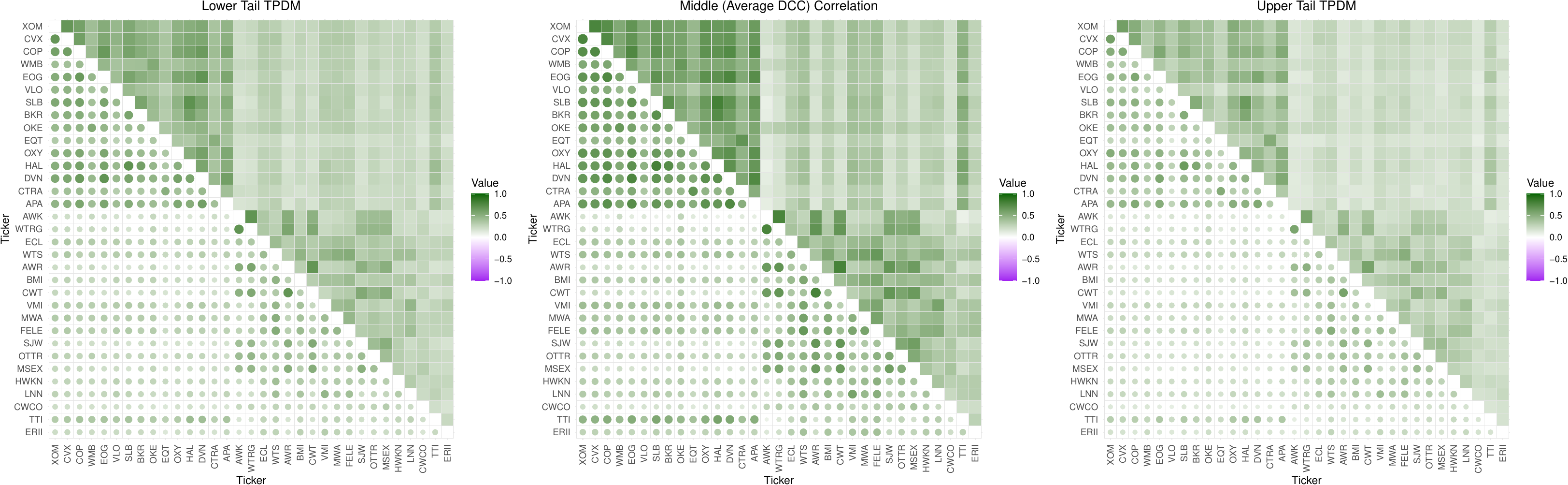}

}

\caption{\label{fig-three-regime-heatmap}Three-panel comparison of the regime-specific correlation / dependence matrices on a common \([-1,1]\) color scale: lower-tail TPDM (left, off-diagonal range {[}0.148, 0.698{]}), time-averaged middle-regime DCC correlation \(\bar R_{\text{mid}}\) (center, off-diagonal range {[}0.085, 0.804{]}), and upper-tail TPDM (right, off-diagonal range {[}0.097, 0.638{]}). A very approximate two-standard deviations above zero is 0.15 for the upper and lower tail, and 0.50 for the middle region.  The high value of the latter is due to the low effective sample size based on the strong persistence of the DCC fitted model. Off-diagonal differences across panels are diagnostic of tail-vs-center and upper-vs-lower asymmetry in the corresponding pair.}

\end{figure}

Figure~\ref{fig-three-regime-heatmap} displays the three regime-specific dependence matrices in a side-by-side arrangement, using a common color scale on the interval \([-1,1]\). The left panel presents the lower-tail TPDM, the central panel reports the DCC-based conditional correlation matrix averaged over the middle subsample, and the right panel depicts the upper-tail TPDM. Because all three panels share an identical color scale, the dependence coefficients are directly comparable across regimes. Exact tail symmetry would be reflected by lower- and upper-tail panels that are nearly mirror images of one another. In contrast, pronounced discrepancies between the central panel and either tail panel highlight stock pairs whose dependence structure is strongly regime-specific.  

\begin{figure}[H]
\centering{
\includegraphics[width=1\linewidth,height=\textheight,keepaspectratio]{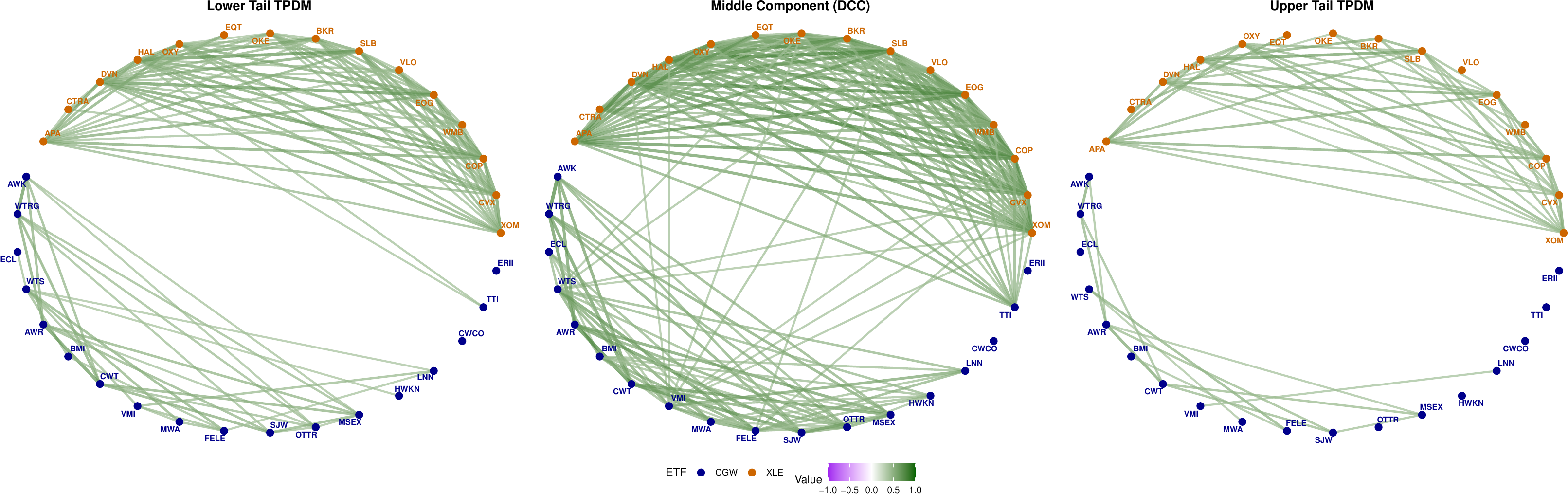}
}
\caption{\label{fig-corr-networks-regime}Correlation networks for the three regime-specific estimators on the full 33-stock cross-section: lower-tail TPDM (left), middle-component DCC time average (center), upper-tail TPDM (right). Nodes are tickers arranged on a circle in the canonical XLE-then-CGW order and colored by ETF membership (XLE dark orange, CGW dark blue). An edge is drawn whenever the absolute value of the matrix entry exceeds the threshold \(0.40\). Edges are colored on the manuscript's diverging purple / white / dark-green scale anchored at zero on \([-1, 1]\), and edge width is proportional to \(|M_{ij}|\). All edges depicted are positive.}
\end{figure}

Complementing the cell-by-cell heatmap in Figure~\ref{fig-three-regime-heatmap}, Figure~\ref{fig-corr-networks-regime} represents the dependence structure as network graphs for the same cross-section of \(33\) stocks across the three regimes. In each panel, the tickers are positioned on a circle, ordered such that the XLE constituents form one contiguous block followed by the CGW constituents. An undirected edge is drawn between two tickers whenever the corresponding pairwise correlation exceeds, in absolute value, a prescribed visual-readability threshold. For all regimes, a common threshold of \(0.40\) is employed.

The choice of the \(0.40\) threshold is guided by the empirical asset-pricing literature on average pairwise equity correlations. For instance, \citet{PolletWilson2010} and \citet{DriessenMaenhoutVilkov2009} document an average realized pairwise correlation of approximately \(28.7\%\) for S\&P 100 constituents over the period 1996–2003. While these studies focus on broad U.S. large-cap universes that are not directly comparable to the industry-focused water- and energy-equity ETF cross-section examined here, they provide a useful quantitative scale against which to interpret the magnitude of correlations in our setting, rather than a strict benchmark for threshold selection.

Figure \ref{fig-corr-networks-regime} shows that the strongest correlations generally occur among assets that belong to the same ETF. A clear exception appears in the Lower Tail TPDM correlation network, where the asset TTI (Tetra Technologies Inc) is highly correlated with two assets in the energy ETF. This is noteworthy because TTI is a water-focused company operating in two segments: Completion Fluids and Products, and Water and Flowback Services. Both business lines are closely linked to the energy sector. TTI is included in CWG because it delivers water management solutions for onshore oil and gas operations \citep{TTI_yahoo_finance}. We would like to highlight this case for the reader because it is one of the few examples of cross-ETF correlation that emerges in the tails.

\begin{figure}[H]
\centering{
\includegraphics[width=0.7\linewidth,height=\textheight,keepaspectratio]{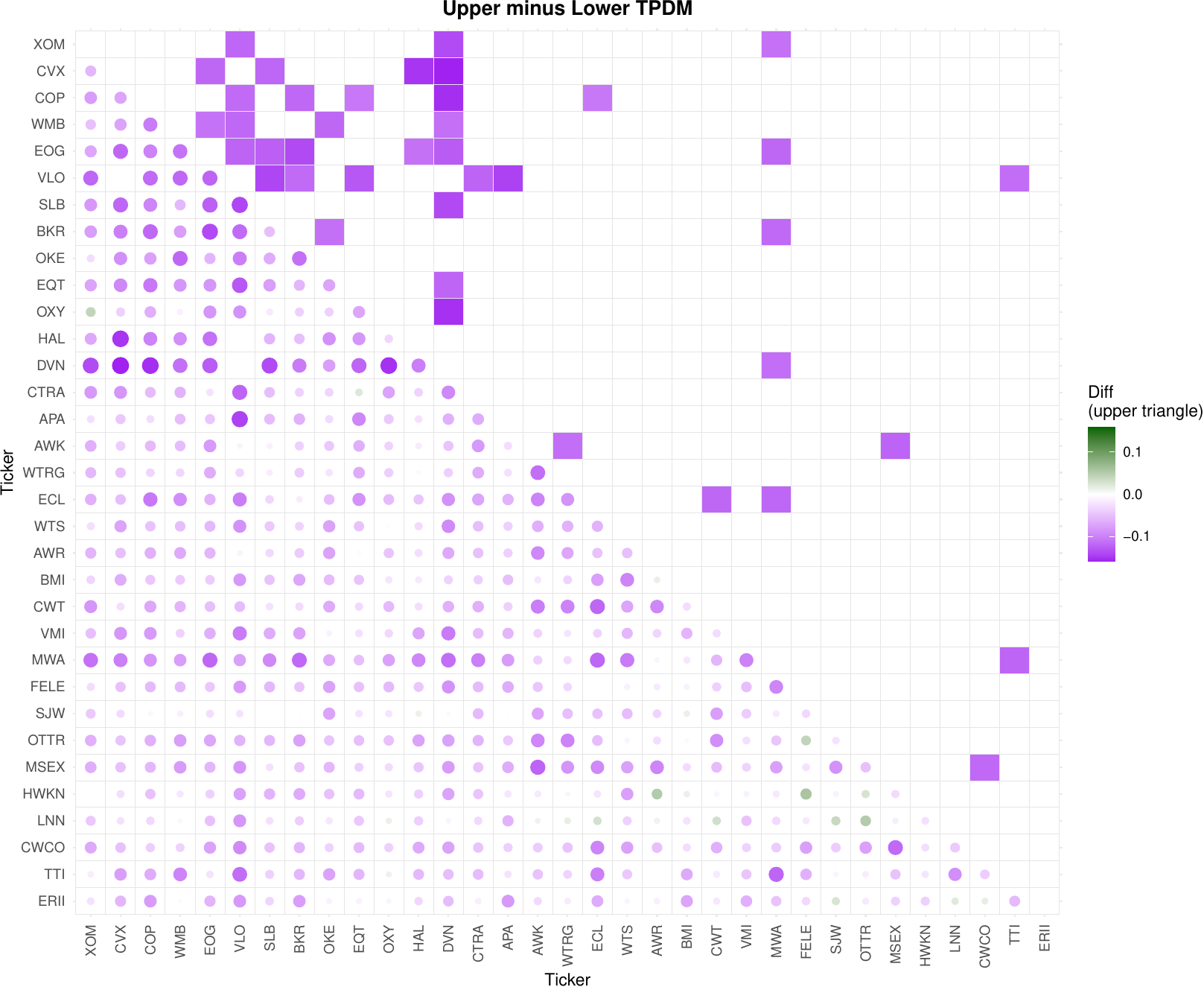}
}

\caption{\label{fig-three-regime-differences}Off-diagonal asymmetry heatmap of \(\widehat\Sigma^{\text{TPD}}_u - \widehat\Sigma^{\text{TPD}}_l\), the difference between the upper-tail and lower-tail TPDMs. The upper triangle carries a symmetric purple / white / dark-green color, applied only to cells whose absolute value exceeds 0.106. Cells inside the noise band are rendered white. The lower triangle shows the difference value as a colored circle on the same diverging scale, with the radius proportional to the absolute difference. The diverging color limit is set to the 90th percentile of the off-diagonal absolute differences, so the largest cells saturate at the scale endpoints to make moderate-magnitude differences visually pronounced.}

\end{figure}

Because the three panels of Figure~\ref{fig-three-regime-heatmap} share a common \([-1, 1]\) color scale dominated by the leading market-mode contribution, relatively small pairwise differences between the lower-tail and upper-tail panels are not visually discernible. To render these differences interpretable, we display the upper-minus-lower TPDM contrast as a separate heatmap in Figure~\ref{fig-three-regime-differences}, employing a narrowly symmetric scale centered at zero. This figure depicts \(\widehat\Sigma^{\text{TPD}}_u - \widehat\Sigma^{\text{TPD}}_l\), which provides a direct measure of tail asymmetry. Entries close to zero indicate approximately symmetric tail co-movements between two assets, whereas entries with large absolute values indicate asset pairs that exhibit substantially stronger dependence in one tail than in the other. The approximate one–standard deviation noise band for each difference is given by \(\sqrt{2/((1-u)T)} \approx 0.106\) at \(u = 0.95\). For the energy ETF, we observe pronounced asymmetry between the upper- and lower-tail dependence measures, with the lower-tail values exceeding the corresponding upper-tail values.

The companion distribution figure below summarizes the entry-wise differences \(R_{ss}\) minus Spearman across all 528 off-diagonal pairs, broken out by sector-pair type pairing a per-group horizontal boxplot with the corresponding density curves. The dashed vertical line marks the point of perfect agreement between the two estimators. Values to the left of the line corresponds to pairs for which \(R_{ss}\) is below the Spearman reference value and values to the right corresponds to the reverse. In general, we see that the range of the difference is quite small, and \(R_{ss}\) are generally below that of the Spearman values. 

\begin{figure}[H]
\centering{
\pandocbounded{\includegraphics[keepaspectratio]{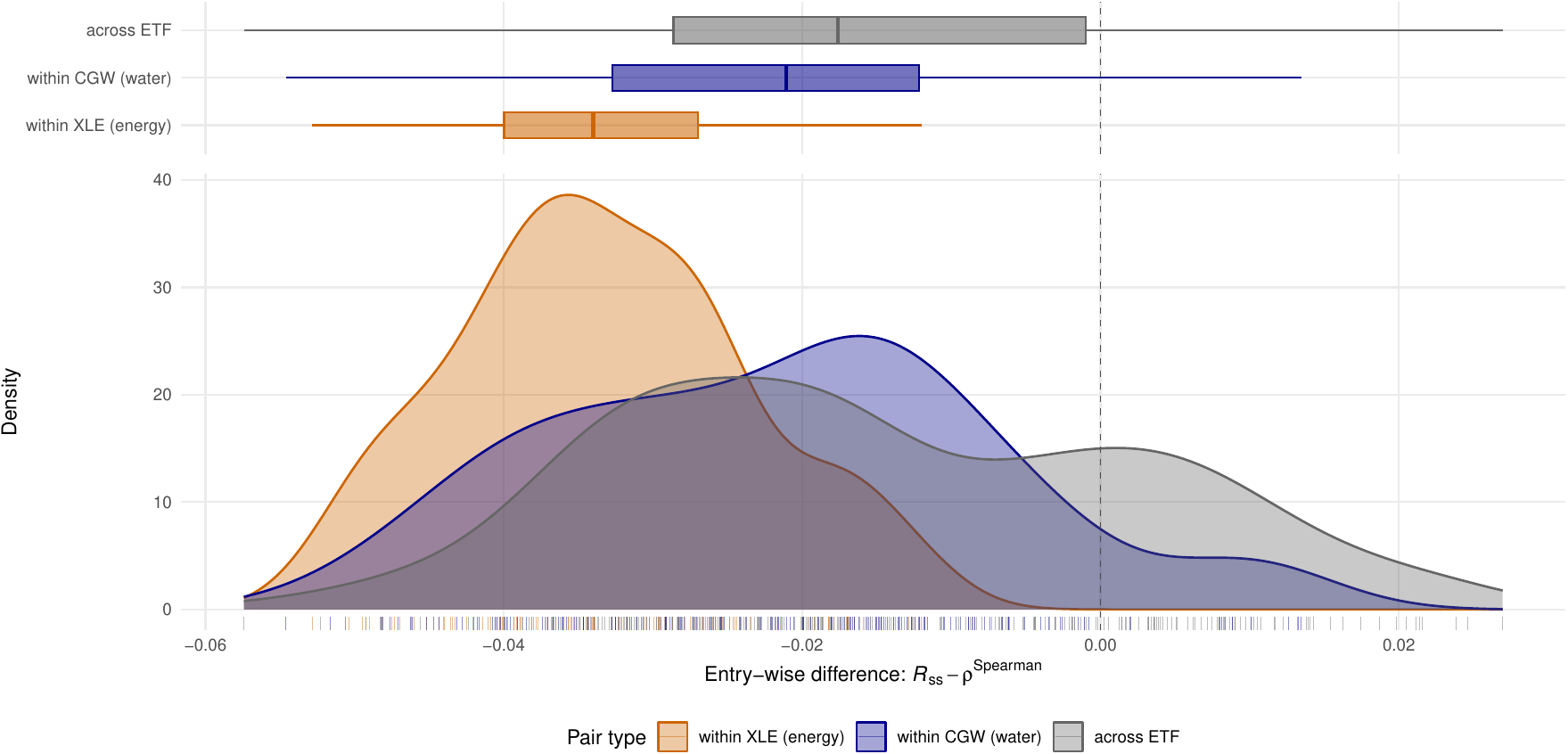}}
}
\caption{\label{fig-ss-minus-sp-distribution}Distribution of the
entry-wise difference \(R_{ss,ij} - \rho^{\text{Spearman}}_{ij}\) across  the 528 off-diagonal pairs in the cross-section.  Top panel: horizontal boxplots per pair type (within XLE energy block,  within CGW water block, across-XLE and CGW pairs), showing the median,  interquartile range, and outliers. Bottom panel: kernel density  estimates of the same three distributions, with a rug below the curves indicating individual pairs. The dashed vertical line is at zero, representing perfect agreement between \(R_{ss}\) and Spearman. This figure  provides the resolved view of the entry-wise differences \(R_{ss}\)  minus Spearman across all off-diagonal pairs, broken out by sector-pair type.}
\end{figure}%

\subsection{A Focus on the Tail Correlation}
To assess whether the upper- and lower-tail dependence matrices differ significantly from zero, we perform a formal comparison with a Gaussian copula estimator applied to the standardized residuals from which the TPDM matrices are constructed. The Gaussian copula provides an adequate representation of dependence in the central part of the distribution. In contrast, statistically significant deviations from the Gaussian-copula null hypothesis of tail independence indicate that the joint tail dependence in this cross-section cannot be captured by any Gaussian copula sharing the same correlation structure in the central regime.

To formalize the pairwise comparison, we implement a Monte Carlo procedure. Across the 528 off-diagonal pairs, the Monte Carlo test rejects the Gaussian copula null hypothesis at the \(5\%\) significance level for 18 upper-tail pairs (3.4\%) and 326 lower-tail pairs (61.7\%) at the manuscript’s primary threshold \(u = 0.95\). At the more extreme threshold \(u = 0.975\), the number of rejections decreases to 11 (2.08\%) for the upper tail and 263 (49.8\%) for the lower tail. Due to the multiple testing we apply the Benjamini-Hochberg FDR-controlled rejection at \(\alpha = 0.05\) at the manuscript's main threshold \(u = 0.95\), yielding 0 rejections in the upper tail, and 261 (49.4\%) for the lower tail. 

The principal substantive finding is the marked asymmetry between the lower and upper tails. Specifically, lower-tail rejection frequencies are approximately an order of magnitude larger than their upper-tail counterparts at both considered thresholds. Further, it is worth noting, that majority of the rejections of independence fall in the cross-market comparisons between XLE and CGW assets in the lower tail. These patterns in dependence are  important when considering within as well as cross-industry risk during market downturns. The observed pattern is consistent with the extensively documented predominance of downside dependence in equity markets and is in line with the directional asymmetry that the regime-aware decomposition was explicitly designed to identify. A nice outcome is the specificity in the equity-to-equity comparisons provided through this formal testing procedure. 

\begin{longtable}[]{@{}
  >{\raggedright\arraybackslash}p{(\linewidth - 10\tabcolsep) * \real{0.0750}}
  >{\raggedright\arraybackslash}p{(\linewidth - 10\tabcolsep) * \real{0.1375}}
  >{\raggedright\arraybackslash}p{(\linewidth - 10\tabcolsep) * \real{0.1500}}
  >{\raggedleft\arraybackslash}p{(\linewidth - 10\tabcolsep) * \real{0.2250}}
  >{\raggedleft\arraybackslash}p{(\linewidth - 10\tabcolsep) * \real{0.2375}}
  >{\raggedleft\arraybackslash}p{(\linewidth - 10\tabcolsep) * \real{0.1750}}@{}}

\caption{\label{tbl-gauss-rejection-by-pair-type}Gaussian-copula null rejection counts broken down by pair type (within-XLE, within-CGW, across-ETF). The first two count columns are per-cell uncorrected rejections at the \(5%
\) level at the two radial-exceedance thresholds \(u = 0.95\) and \(u = 0.975\); the third is the Benjamini-Hochberg FDR-controlled rejection count at \(\alpha = 0.05\) at the manuscript's main threshold \(u = 0.95\), with the BH correction applied within each tail across the 528 unique off-diagonal pairs. The BH-FDR \(\alpha\) is matched to the per-cell \(\alpha = 0.05\) used in the first columns. }

\tabularnewline

\toprule\noalign{}
\begin{minipage}[b]{\linewidth}\raggedright
Tail
\end{minipage} & \begin{minipage}[b]{\linewidth}\raggedright
Pair type
\end{minipage} & \begin{minipage}[b]{\linewidth}\raggedright
Total pairs
\end{minipage} & \begin{minipage}[b]{\linewidth}\raggedleft
Rejections @ 0.95
\end{minipage} & \begin{minipage}[b]{\linewidth}\raggedleft
Rejections @ 0.975
\end{minipage} & \begin{minipage}[b]{\linewidth}\raggedleft
BH-FDR @ 0.95
\end{minipage} \\
\midrule\noalign{}
\endhead
\bottomrule\noalign{}
\endlastfoot
Upper & within XLE & 105 & 0 (0.0\%) & 0 (0.0\%) & 0 (0.0\%) \\
Upper & within CGW & 153 & 15 (9.8\%) & 6 (3.9\%) & 0 (0.0\%) \\
Upper & across ETF & 270 & 3 (1.1\%) & 5 (1.9\%) & 0 (0.0\%) \\
Lower & within XLE & 105 & 70 (66.7\%) & 47 (44.8\%) & 63 (60.0\%) \\
Lower & within CGW & 153 & 91 (59.5\%) & 65 (42.5\%) & 75 (49.0\%) \\
Lower & across ETF & 270 & 165 (61.1\%) & 151 (55.9\%) & 123 (45.6\%) \\

\end{longtable}

To localize where the Gaussian-copula rejection is concentrated within the cross-section, we visualize the per-pair standardized rejection statistic \(z_{ij} = (\widehat\Sigma^{\text{TPD}}_{ij} - \mu^{\text{null}}_{ij}) / \sigma^{\text{null}}_{ij}\) as a pair of heatmaps. Here \(\mu^{\text{null}}_{ij}\) and \(\sigma^{\text{null}}_{ij}\) are the mean and standard deviation of the off-diagonal entry \((i,j)\) across the \(M\) Monte Carlo replicates. Large positive \(z\) indicates strong rejection of the Gaussian-copula null in favor of higher observed tail dependence than the Gaussian copula predicts. Values close to zero indicate pairs whose observed tail dependence sits inside the Gaussian-copula null distribution and so cannot be distinguished from it on this sample. Negative values are possible in principle but very rare in equity panels, and they would correspond to pairs whose observed tail dependence is weaker than the Gaussian-copula prediction. We do go ahead and perform the Benjamini-Hochberg multiple testing adjustment (BH-FDR). At a 0.10 level of significance, the BH-FDR finds 358 out of 528 off-diagonal pair correlations significant in the lower tail. However, in the upper tail no pair correlations are signficantly different from zero. 

\begin{figure}[H]

\centering{

\includegraphics[width=1\linewidth,height=\textheight,keepaspectratio]{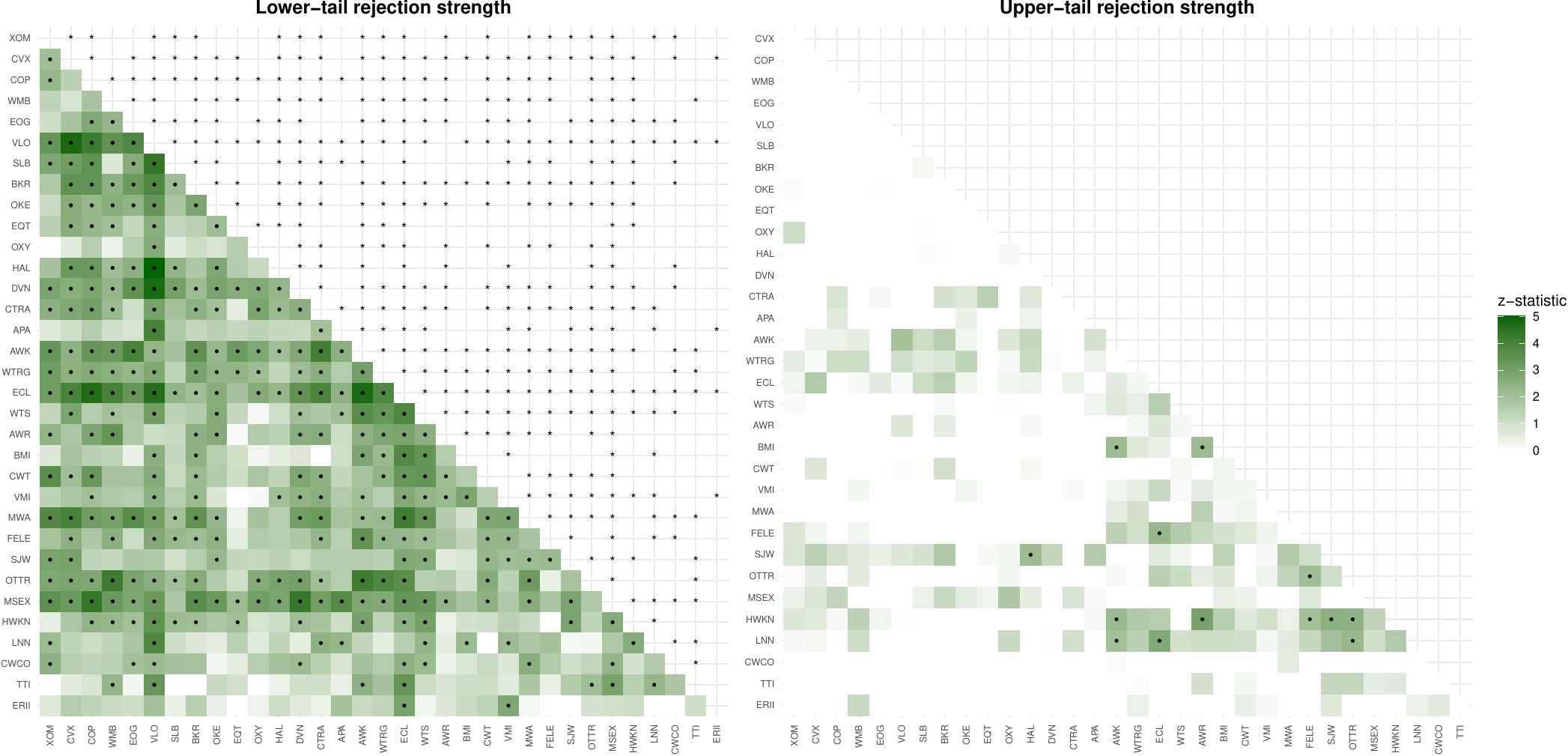}

}

\caption{\label{fig-gauss-rejection-strength}Per-pair standardized rejection statistic \(z_{ij} = (\widehat\Sigma^{\text{TPD}}_{ij} - \mu^{\text{null}}_{ij}) / \sigma^{\text{null}}_{ij}\) at the main threshold \(u = 0.95\). The TPDM is symmetric, so each panel uses its lower triangle to display color (plus a per-cell marker) and its upper triangle to display the BH-FDR significance marker. Lower-triangle cells are colored on a white-to-dark-green sequential scale shared between both panels, with the upper limit set to \(\max_{ij} z_{ij}\) across both tails; negative \(z\) values (pairs whose observed tail co-exceedance falls below the Gaussian-copula null mean, a one-sided non-rejection) are clamped to white. A small black dot is overlaid on any lower-triangle cell with \(z_{ij} > 1.96\) (per-cell uncorrected normal-approximation cutoff). Upper-triangle cells carry a black asterisk for pairs whose one-sided Monte Carlo \(p\)-value, computed as the fraction of MC replicates with simulated TPDM entry at least the observed value, is significant at the \(10\%\) level after Benjamini-Hochberg FDR control across the 528 unique off-diagonal pairs within that tail. Diagonals are blanked. Left: lower-tail TPDM. Right: upper-tail TPDM.}

\end{figure}%

The rejection counts close the loop with our claim that the Gaussian-copula assumption breaks down precisely on the joint dependence that drives the most consequential risk metrics. Pairs that reject the Gaussian-copula null in either tail are pairs for which a parametric Gaussian-copula model would systematically under-state joint tail co-movement, and the per-pair-type breakdown above localizes whether the rejection is concentrated within ETF (water or energy individually) or across ETF. The headline asymmetry is that the lower-tail rejection rate (63.8\% of off-diagonal pairs at \(u = 0.95\)) is roughly 18 times higher than the upper-tail rejection rate (3.4\%). This is exactly the downside-vs-upside dependence asymmetry the regime-aware decomposition was designed to capture and the Gaussian copula is incapable of producing. 

\subsection{Computational Cost}\label{computational-cost}

The pipeline assembled here sits at a different point on the cost-flexibility frontier from the Gaussian copula benchmark that dominates operational practice. The Gaussian copula's fit is one \(\mathcal O(N^{2} T)\) pass to form the sample correlation of the normal scores plus one \(\mathcal O(N^{3})\) Cholesky of that \(N \times N\) matrix. After that, every density evaluation, conditional, and simulated draw reduces to solving a triangular system, with no time-indexed multivariate factorization. The regime-aware pipeline trades that simplicity for explicit modeling of the three regimes, paying three separable costs.

\begin{itemize}
  \item \textbf{Univariate GARCH(1,1) filter, $\mathcal O(NT)$.} Each of the $N$ series is fit independently by quasi-maximized likelihood on its own conditional-variance recursion. The recursion is one forward pass through length-$T$ data, so each likelihood evaluation is $\mathcal O(T)$; with a bounded number of optimizer iterations the per-series cost is $\mathcal O(T)$, giving $\mathcal O(NT)$ across the cross-section.
  \item \textbf{Two-stage DCC on the standardized residuals, $\mathcal O(N^{3} T)$ per fit.} The dominant per-step cost is the Cholesky factorization of the $N \times N$ conditional correlation matrix $R_{t}$, which is $\mathcal O(N^{3}/3)$ flops. The log-determinant and quadratic form fall out of that factorization, and the $Q_{t}$ recursion update is the cheaper $\mathcal O(N^{2})$. Over $T$ time steps each likelihood evaluation is $\mathcal O(N^{3} T)$, and because the DCC has only two free parameters $(a, b)$ the optimizer converges in a bounded number of likelihood calls.
  \item \textbf{Tail TPDM, $\mathcal O(N^{2} |\mathcal T_{\text{tail}}|)$ per tail.} The TPDM has $N(N-1)/2$ unique off-diagonal entries; each is an angular second moment computed over the pair's radial-exceedance set, which has expected size $|\mathcal T_{\text{tail}}| \approx (1 - u)\, T$ at threshold $u$. The marginal ECDF / Fr\'echet transform pass adds a subdominant $\mathcal O(NT)$.
\end{itemize}

On the cross-section studied here (\(N = 33\), \(T = 3,587\)), the total cost is dominated by the DCC's \(N^{3} T\) multivariate factorization. The fit remains tractable on a reasonably powered laptop or single workstation. For substantially larger panels the high-dimensional methods of \citet{YaoZouXing2024} and \citet{TongHansenArchakov2025} provide natural scaling routes.

\section{Discussion and Next Steps}\label{discussion-and-next-steps}

This manuscript underscores the critical importance of explicitly accounting for both upside and downside tail behavior when analyzing dependence structures and risk measures. The highly significant and numerous marginal correlations identified in the lower regime—both within individual ETF and across ETF—provide compelling evidence that downside risk exhibits dynamics that are distinct from those observed in upside regimes and in more tranquil market conditions. It is therefore essential that methodological frameworks explicitly incorporate this asymmetry. We propose an approach that does so directly, enabling a clear and systematic characterization of the heterogeneous impact of dependence across all three regimes. 

A second robustness check, recommended for applied implementations, concerns the choice of parameterization for the DCC component. A key enhancement of the standard DCC model is the corrected DCC (cDCC) of \citet{Aielli2013}. In the conventional specification, the recursion
\[
Q_t = (1 - a - b)\,\bar Q + a\, z_{t-1} z_{t-1}^{\top} + b\, Q_{t-1}
\]
is estimated using the sample correlation matrix of the standardized residuals as a plug-in estimator for \(\bar Q\). However, the marginal process \(z_t\) satisfies \(E[z_t z_t^{\top}] = E[R_t]\) rather than \(E[Q_t]\). Since the standardization
\[
R_t = \operatorname{diag}(Q_t)^{-1/2}\, Q_t\, \operatorname{diag}(Q_t)^{-1/2}
\]
is a nonlinear transformation, these unconditional moments do not coincide in general. Consequently, the standard two-step QMLE for \((a, b)\) is, in general, inconsistent for its own population target. Aielli’s cDCC specification resolves this issue by reparameterizing the recursion in terms of
\[
z^{*}_t = \operatorname{diag}(Q_t)^{1/2}\, z_t,
\]
whose unconditional second moment \(\bar S = E[z^{*}_t (z^{*}_t)^{\top}]\) is consistently estimable via its sample analogue, thereby restoring the consistency of the two-step QMLE under standard regularity conditions. \citet{CaporinMcAleer2013} provide a comprehensive review of the existing results and outstanding issues in this literature. The \texttt{rmgarch} package, on which the remainder of our empirical pipeline relies, does not currently implement Aielli’s reparameterization as a built-in option. To streamline the computations for this study, we therefore restricted attention to the functionality available within \texttt{rmgarch}. For formal implementation in empirical applications, however, we recommend adopting a workflow in which standard cDCC estimation replaces the DCC specification employed here.

The TPDM methodology introduced in this paper yields marginal tail-dependence summaries, that is, each off-diagonal entry of \(\widehat\Sigma^{\text{TPD}}\) quantifies the pairwise extremal co-movement between two assets without conditioning on the remaining components of the cross-section. This is the natural analogue, in an extreme-value context, of working with the marginal correlation matrix in a Gaussian setting. However, by analogy with classical multivariate time-series analysis and regression, it is the partial correlation matrix, given by the off-diagonal elements of the inverse covariance matrix, that forms the basis of modern graphical-model theory, conditional-independence structure learning, and sparsity-inducing procedures such as the graphical lasso and its extensions. It is therefore a natural progression for our framework to move from a marginal TPDM representation to an analog that encodes conditional extremal independence.

Two recent contributions provide the methodological basis for such an extension. \citet{GongEtAl2024} define a partial tail-correlation coefficient (PTCC) by extending the usual partial-correlation construction to the transformed-linear algebra on the positive orthant that underlies the Cooley–Thibaud framework. This construction enables testing whether the tails of a pair of variables are partially uncorrelated, conditional on the remainder of the cross-section. Moreover, \citet{GongEtAl2024} demonstrate how the PTCC can be integrated with high-dimensional estimation techniques, including the graphical lasso with Laplacian spectral constraints, to directly recover an extremal network structure. In parallel, \citet{KimLee2024} develop a complementary hypothesis-testing framework for partial tail correlation that provides a formal inferential mechanism for deciding whether edges are absent or present in such extremal graphical models.

Future research will also address sparse multivariate extreme-value modeling, wherein the estimated graph provides the conditional-independence skeleton on which a Hüsler–Reiss model \citep{HuslerReiss1989} or a related parametric extreme-value family can be fitted \citep{EngelkeEtAl2024}. Incorporating this framework into our current analysis pipeline would only necessitate replacing the TPDM step with the PTCC-and-thresholding procedure of \citet{GongEtAl2024}. All preceding components, including univariate GARCH modeling, middle-region filtering, and DCC estimation, as well as all subsequent applications of the resulting dependence summaries, would remain unchanged.

\section{AI Usage:} The authors used Claude for Mac 1.12603.1 (Anthropic) to assist in writing reproducible code. Code was designed, reviewed, tested, and validated by the authors ensuring the intended methodology is implemented. The authors take full responsibility for the correctness of the code, the integrity of the underlying data, and the accuracy of all reported results. Minor writing assistance was supported by Writefull through Overleaf. Neither Claude nor Writefull were allowed to train on the materials.

\section{Acknowledgement:} The authors acknowledge the Center for Computational Finance and Economic Systems (CoFES) (see cofes.rice.edu), of which Ensor is director. The vision for the paper materialized from the ASA Summer 2026 Conference NSF@75 Advancing Statistical Science for a Data-Driven World and the paper was first presented at STATFIN2026 (see statfin.cmi.ac.in/ISBIS2026). The authors thank the participants for their useful comments. 

\section{References}\label{references}

\renewcommand{\bibsection}{}
\bibliography{refs.bib}

\end{document}